\documentclass[pre,twocolumn,showpacs,showkeys,preprintnumbers,floatfix]{revtex4-1}

\usepackage{etex}
\usepackage{ifpdf}
\usepackage{hyperref}
\usepackage{dcolumn}
\usepackage{url}
\usepackage{amsmath}
\usepackage{amscd}
\usepackage{amsfonts}
\usepackage{amssymb}
\usepackage{amsthm}
\usepackage{bigints}
\usepackage{xfrac}
\usepackage{braket}
\usepackage{bm}   
\usepackage{bbm}
\usepackage{verbatim}
\usepackage{stmaryrd}
\usepackage{xcolor}
\usepackage{setspace}
\usepackage{tikz}
\usetikzlibrary{arrows}
\usetikzlibrary{automata}
\usetikzlibrary{positioning}
\usetikzlibrary{plotmarks}
\usepackage{pgfplots}
\pgfplotsset{compat=newest}
\usepackage{hyphenat}
\usepackage{hyperref}
\usepackage{mathrsfs}

\newcommand{\cs}{\causalstate}
\newcommand{\Abet}{\ProcessAlphabet}
\newcommand{\MS}{\MeasSymbol}

\newcommand{\ms}{\meassymbol}
\newcommand{\SSet}{\CausalStateSet}
\newcommand{\St}{\CausalState}
\newcommand{\st}{\causalstate}

\renewcommand{\H}{\operatorname{H}}

\definecolor{commentred}{rgb}{0.8, 0.2, 0.2}
\definecolor{commentgreen}{rgb}{0.2, 0.8, 0.2}

\renewcommand{\H}{\operatorname{H}}

\newcommand{\qML} {q-machine}
\newcommand{\qMLs} {q-machines}

\newcommand{\QML} {q-Machine}

\newcommand{\synk}{\mathrm{SINK}} 

\newcommand{\dGL}{(\delta G)^{(L)}}

\usepackage{amsmath,amssymb,amsthm,mathrsfs,amsfonts,dsfont}

\tikzstyle{vaucanson}=[
  node distance=2.5cm,
  bend angle=15,
  auto,
  every loop/.style={},
  every edge/.style={->,draw=black,line width=1.2,>=latex,shorten <=1pt,shorten >=1pt},
  every state/.style={draw=black,line width=2,font=\large},
  loop right/.style={right,out=22,in=-22,loop},
  loop above/.style={above,out=112,in=68,loop},
  loop left/.style={left,out=202,in=158,loop},
  loop below/.style={below,out=292,in=248,loop},
  loop above right/.style={right,out=67,in=23,loop},
  loop above left/.style={left,out=157,in=113,loop},
  loop below left/.style={left,out=247,in=203,loop},
  loop below right/.style={right,out=337,in=293,loop},
]

\theoremstyle{plain}    \newtheorem{Lem}{Lemma}
\theoremstyle{plain}    \newtheorem*{ProLem}{Proof}
\theoremstyle{plain}    \newtheorem{Cor}{Corollary}
\theoremstyle{plain}    \newtheorem*{ProCor}{Proof}
\theoremstyle{plain}    \newtheorem{The}{Theorem}
\theoremstyle{plain}    \newtheorem*{ProThe}{Proof}
\theoremstyle{plain}    \newtheorem{Prop}{Proposition}
\theoremstyle{plain}    \newtheorem*{ProProp}{Proof}
\theoremstyle{plain}    
\theoremstyle{plain}    
\theoremstyle{plain}    \newtheorem{Def}{Definition}
\theoremstyle{plain}    
\theoremstyle{plain}

\DeclareMathOperator*{\argmax}{argmax}

\newcommand{\eM}     {\mbox{$\epsilon$-machine}}
\newcommand{\eMs}    {\mbox{$\epsilon$-machines}}
\newcommand{\EM}     {\mbox{$\epsilon$-Machine}}

\newcommand{\MeasAlphabet}  {\mathcal{A}}
\newcommand{\MeasSymbol}   { {X} }
\newcommand{\meassymbol}   { {x} }
\newcommand{\MeasSymbols}[2]{ \MeasSymbol_{#1:#2} }
\newcommand{\meassymbols}[2] { \meassymbol_{#1:#2} }

\newcommand{\Future} { \MeasSymbols{0}{} }
\newcommand{\future} { \meassymbols{0}{} }

\newcommand{\CausalState}   { \mathcal{S} }

\newcommand{\causalstate}   { \sigma }
\newcommand{\CausalStateSet}    { \boldsymbol{\CausalState} }

\newcommand{\Prob}      {\Pr} 

\newcommand{\Cmu}       {C_\mu}

\newcommand{\EE}        {{\bf E}}

\newcommand{\PC}        {\chi}

\newcommand{\ProcessAlphabet}   {\MeasAlphabet}

\newcommand{\forward}{+}
\newcommand{\reverse}{-}
\newcommand{\forwardreverse}{\pm} 

\newcommand{\FutureCausalState} { {\CausalState}^{\forward} }

\newcommand{\PastCausalState}   { {\CausalState}^{\reverse} }

\newcommand{\lastindex}[2]{
  \edef\tempa{0}
  \edef\tempb{#2}
  \ifx\tempa\tempb
    \edef\tempc{#1}
  \else
    \edef\tempa{0}
    \edef\tempb{#1}
    \ifx\tempa\tempb
      \edef\tempc{#2}
    \else
      \edef\tempc{#1+#2}
    \fi
  \fi
  \tempc
}

\newcommand{\CSjoint}[1][,]{
   \edef\tempa{:}
   \edef\tempb{#1}
   \ifx\tempa\tempb
      \ensuremath{\FutureCausalState\!#1\PastCausalState}
   \else
      \ensuremath{\FutureCausalState#1\PastCausalState}
   \fi
}

\newif\ifpm
\edef\tempa{\forwardreverse}
\edef\tempb{\pm}
\ifx\tempa\tempb
   \pmfalse
\else
   \pmtrue
\fi

\colorlet {R_color}    {blue}
\colorlet {k_color}    {black!30!green}

\def\clap#1{\hbox to 0pt{\hss#1\hss}}

\begin{document}

\title{A Closed-Form Shave from Occam's Quantum Razor:\\Exact Results for Quantum Compression}

\author{Paul M. Riechers}
\email{pmriechers@ucdavis.edu}
\author{John R. Mahoney}
\email{jrmahoney@ucdavis.edu}
\author{Cina Aghamohammadi}
\email{caghamohammadi@ucdavis.edu}
\author{James P. Crutchfield}
\email{chaos@ucdavis.edu}

\affiliation{Complexity Sciences Center and Department of Physics\\
University of California at Davis\\
One Shields Avenue, Davis, CA 95616}

\date{\today}
\bibliographystyle{unsrt}
 
\begin{abstract}
The causal structure of a stochastic process can be more efficiently
transmitted via a quantum channel than a classical one, an advantage that
increases with codeword length. While previously difficult to compute, we
express the quantum advantage in closed form using spectral decomposition,
leading to direct computation of the quantum communication cost at all encoding
lengths, including infinite. This makes clear how finite-codeword compression
is controlled by the classical process' cryptic order and allows us to analyze
structure within the length-asymptotic regime of infinite-cryptic order (and
infinite Markov order) processes.
\end{abstract}


\preprint{Santa Fe Institute Working Paper 15-10-XXX}
\preprint{arxiv.org:1510.XXXX [physics.gen-ph]}

\maketitle



\setstretch{1.1}


\section{Introduction}

Defining and quantifying ``pattern'' and ``structure'' have been
active endeavors for decades \cite{Crut12a,Ball99a,Hoyl06a,Kant06a}. They are
especially important and challenging when it comes to the complex patterns
spontaneously generated by nonlinear dynamical systems and hidden stochastic
processes. These studies seek to answer questions such as, ``how unpredictable
is this pattern?'' and ``how much memory is needed to accurately predict its
next element?'' Computational mechanics \cite{Crut88a,Shal98a}, an
extension of statistical mechanics, was established to answer these
very questions. In much of computational mechanics, the emphasis is on
stationary stochastic processes. Here, we explore how their internal structure
generates patterns.

The question of a process's internal structure naturally arises for two
observers, Alice and Bob, who wish to efficiently synchronize their predictions
of a given process over a classical communication channel. What is the minimal
amount of information they that must communicate? The answer is given by the
process' statistical complexity $\Cmu$ \cite{Crut12a}.

A closely related question immediately suggests itself: is it more efficient to
synchronize via a quantum communication channel? Extending early answers
\cite{Gu12a,Gmein11a}, we recently introduced a sequence of constructions
(\qMLs) that offer substantially improved quantum compression \cite{Maho15a}.
Each codeword length $L$ yields a quantum communication rate $C_q(L)$, where
$C_q(L) \leq \Cmu$. Moreover, we showed that maximum compression $C_q(\infty) =
C_q(k)$ is achieved at a codeword length called the \emph{cryptic order} $k$
\cite{Crut08a}---a recently discovered classical, topological property that is
a cousin to the more familiar stochastic process Markov order.

The following develops the analytical underpinnings for the \qML\ construction
and for computing the quantum communication costs $C_q(L)$ \cite{Maho15a}.
We present a closed-form
expression for the overlaps between quantum signal states---overlaps that lead
to quantum compressibility. Overlaps are expressed in terms of the spectrum
(and related projection operators) of the quantum pairwise-merger machine,
introduced here. This leads us to a decomposition of the quantum compression 
into two qualitatively distinct parts: a finite-horizon contribution
at any cryptic order and a persistent infinite-horizon contribution that arises
only for infinite-cryptic processes.

An even greater advantage obtains when using a surrogate for the quantum
density matrix---a fixed-size Gram matrix, which has the same spectrum.
Critically, the surrogate matrix is linear in the overlaps and is
straightforward to calculate. Using it, the quantum communication costs
$C_q(L)$ (and $C_q(\infty)$) are calculable from a closed-form expression, in
some cases analytically and, in all cases, with an unprecedented level of
numerical accuracy and efficiency. Moreover, the generic behavior of $C_q(L)$
for large $L$ and infinite cryptic order can be probed. One lesson, for
example, is that optimal quantum compression is achieved asymptotically, but
with exponentially diminishing returns. And, we discover and explain why the
quantum advantage oscillates with increasing codeword length: the oscillations
reflect the structure of a process' crypticity---how its internal state information,
hidden even upon all subsequent observations, reveals itself only slowly
with the additional information of past observations provided with
increasing codeword length.

The next section reviews the causal-state representation that computational
mechanics employs for stochastic processes---the \eM---and recalls the new
representation---the \qML---appropriate to quantum communication. We then show
how to build quantum codewords out of a classical process' causal states and
observed sequences and how to monitor the codewords' quantum overlap. Central to
determining the proper overlap structure, we introduce the quantum
pairwise-merger machine (QPMM), detailing an algorithm for its construction and
illustrating its use via several example processes. Next, we turn to measuring
quantum compressibility via the von Neumann entropy of the overlap density
matrix. Spectral decomposition of the QPMM transition dynamic, though, leads to
closed-form expressions for the elements of the Gram matrix, which yields the
quantum coding cost at any length. We explore several example processes with
these tools to illustrate a number of features when quantum compressing
structured classical processes. These include finite and infinite cryptic-order
processes and the universal behavior of quantum compression in the asymptotic
codeword-length limit. We conclude with a broader discussion of the application
of these results and brief comments on how they affect our view of the
mechanisms by which classical and quantum processes generate the patterns we
see in the physical world.

\section{Two Representations of a Process}

The objects of interest are discrete-valued, stationary, stochastic processes
generated by finite hidden Markov models (HMMs). In particular, we consider
edge-output HMMs (or Mealy HMMs) where the observed symbol is generated upon
transition between states. Rather than focus on generating models, more
prosaically we can also think of a \emph{process} consisting of a bi-infinite
sequence $\MS_{-\infty:\infty} = \ldots \MS_{-2} \MS_{-1} \MS_0 \MS_1 \MS_2
\ldots$ of random variables $\MS_t$ that take on one or another value in a
discrete alphabet: $\ms_t \in \Abet$. A process' \emph{language} is that set of
words $w = \ms_0 \ldots \ms_{L-1}$ of any length $L$ generated with
positive probability. We consider two representations of a given process, first
a canonical classical representation and then a newer quantum representation.

\subsection{\EM }

While a given process generally has many alternative HMM representations, there
exists a unique, canonical form---the process's \emph{\eM} \cite{Crut12a}. An
equivalence relation applied to the random variable sequence
$\MS_{-\infty:\infty}$ defines the process' causal states, which encapsulate
all about individual pasts relevant for predicting the future. Said
another way, causal states are the minimal sufficient statistic of the past
$\MS_{\infty:0}$ for predicting the future $\MS_{0:\infty}$. (Note that we use
array indexing that is left inclusive, but right exclusive.)

\begin{Def}
A process' \emph{\eM} $\mathcal{M}$ is the tuple
$\big( \CausalStateSet , \, \MeasAlphabet, \, \{ T^{(\ms)} \}_{ \ms \in \MeasAlphabet} , \, \boldsymbol{\pi}  \big)$.
\end{Def}

$\CausalStateSet$ is the set $\{ \cs_0, \cs_1, \ldots \}$ of the process' causal states,
$\Abet$ is the set of possible symbol outputs $\ms$, $\bigl\{
T^{(\ms)} : T^{(\ms)}_{i,j} = \Pr(\cs_j, \ms | \cs_i) \bigr\}_{\ms \in \Abet}$ are the labeled transition matrices, and
$\boldsymbol{\pi}$ the stationary distribution over states. The probability that a
word $w$ is generated by the \eM\ is given in terms of the labeled transition
matrices and the initial state distribution:
\begin{align*}
\Prob(w) = \boldsymbol{\pi} \prod_{i=0}^{L-1} T^{(\ms_i)}\mathds{1},
\end{align*}
where $\mathds{1} = [1,\ldots, 1]^\top$. When these probabilities are constructed
to agree with those of the words in a given process language, the \eM\ is said
to \emph{generate} or \emph{represent} that process.

The ensemble temporal evolution of internal state probability $\boldsymbol{\mu} = \left(
\mu_0, \ldots, \mu_{|\SSet|-1} \right)$, with $\mu_i = \Pr(\cs_i)$, is given by:
\begin{align}
\boldsymbol{\mu}(t+1) = \boldsymbol{\mu}(t) T
  ~,
\label{eq:StateTransitionOp}
\end{align}
where the net transition matrix $T$ is the sum over all output symbols:
\begin{align}
T := \sum_{\ms \in \MeasAlphabet}T^{(\ms)}
  ~.
\label{eq:StateTransitionOp}
\end{align}
Transition probabilities are normalized. That is, the transition matrix $T$
is \emph{row-stochastic}:
\begin{align*}
\sum_{j=1}^{|\SSet|} T_{i,j} = \sum_{j=1}^{|\SSet|} \sum_{\ms \in \MeasAlphabet} \Prob(\cs_j, \ms \vert \cs_i) = 1
  ~.
\end{align*}
Its component matrices $T^{(\ms)}_{ij}$ are said to be \emph{substochastic}.
Under suitable conditions on the transition matrix, $\lim_{t \to \infty} \boldsymbol{\mu} (t) =
\boldsymbol{\pi}$.

\emph{Unifilarity}, an inherent property of \eMs, means that for each state
$\cs_i$, each symbol $x$ may lead to at most one successor state $\cs_j$
\cite{Ash65a}. In terms of the labeled transition matrices, for each row $i$
and each symbol $x$ the row $T^{(\ms)}_{ij}$ has at most one nonzero entry.
This property falls naturally out of the equivalence relation used to define
causal states \cite{Crut12a}. We also will have occasion to speak of a
\emph{counifilar} HMM, which is the analogous requirement of unique labeling on
transitions coming \emph{into} each state.

One of the most important complexity measures for a process is its statistical
complexity $\Cmu$ \cite{Crut12a}. Used in a variety of contexts, it quantifies a process' minimal description size. 
\begin{Def}
The \emph{statistical complexity} $\Cmu$ of a process is the Shannon entropy of
the stationary distribution over its \eM's causal states:
\begin{align*}
\Cmu & = H[\boldsymbol{\pi}] \\
     & = -\sum_{i = 1}^{|\SSet|} \pi_i \log \pi_i
	 ~.
\end{align*}
\end{Def}
The statistical complexity has several operational meanings. For example, it is
the average amount of information one gains on determining in which causal
state a process is. It is also the minimal amount of information that must be
stored from the past to optimally predict the future. Most pertinent to our
purposes here, it also quantifies the communication cost of synchronizing two
predicting agents through a classical channel \cite{Maho15a}.

\subsection{\QML\ }

The \qML\ is a representation of a classical process that makes use of quantum
mechanics. Introduced in Ref. \cite{Maho15a}, it offers the process' most
complete quantum compression known so far.

A process' \qML\ is constructed by first selecting a \emph{codeword length}
$L$. The \qML\ (at $L$) consists of a set $\{ \ket{\eta_i(L)}
\}_{i=1}^{|\SSet|}$ of pure quantum \emph{signal states} that are in
one-to-one correspondence with the classical causal states $\st_i \in
\CausalStateSet$. Each signal state $\ket{\eta_i(L)}$ encodes the set of
length-$L$ words $\{w: \Pr(w|\st_i) > 0\}$ that may follow causal state
$\st_i$, as well as the corresponding conditional probability:
\begin{align}\label{eq:qmdef}
\ket{\eta_i(L)} \equiv
  \sum \limits_{w \in \Abet^L}
  \sum \limits_{\st_j \in \CausalStateSet}
  {\sqrt{\Prob(w, \st_j | \st_i)}
  ~ \ket{w} \ket{\st_j}}
  ~,
\end{align}
where $\big\{\ket{w}\big\}_{w \in \MeasAlphabet^L}$ denotes an orthonormal
basis in the ``word'' Hilbert space with one dimension for each possible word
$w$ of length $L$. Similarly, $\big\{\ket{\st_j}\big\}_{j=1}^{|\SSet|}$ denotes
an orthonormal basis in the ``state'' Hilbert space with one dimension for each
classical causal state. The ensemble of length-$L$ quantum signal states is
then described by the density matrix:
\begin{align}\label{denmat}
\rho(L) = \sum_{i=1}^{|\SSet|} {\pi_i \ket{\eta_i(L)} \bra{\eta_i(L)}}
  ~.
\end{align}

The ensemble's von Neumann entropy (VNE) is defined in terms of its density
matrix: $S(\rho) = - \mathrm{tr}[\rho \log(\rho)]$, where tr$[\cdot]$ is the
trace of its argument. Paralleling the classical statistical complexity,
the quantity:
\begin{align}
C_q(L) & \equiv S(\rho(L)) \nonumber \\
       & = - \mathrm{tr} \bigl[ \rho(L) \log(\rho(L)) \bigr]
\end{align}
has the analogous operational meaning of the communication cost over a
quantum channel. This
is of interest since it is generically smaller \cite{Maho15a}: $C_q(L) \leq
\Cmu$. In fact, $\Cmu = C_q$ if and only if the process' \eM\ is
counifilar---there are no states with (at least) two similarly labeled incoming
edges \cite{Gu12a}. Notably, processes with counifilar \eMs\ are a vanishing
proportion of all processes, as we increase state size \cite{Jame10a}. The
consequence is that almost all classical processes can be more compactly
represented using quantum mechanics. This presents an opportunity to use
quantum encoding to more efficiently represent a process.

Actually computing the quantum cost $C_q(L)$ is challenging, however.
Circumventing this difficulty is one of the main motivations for the following
developments. The eventual results, though, allow us to move beyond this
practical concern to a deeper appreciation of quantum structural complexity.

\section{Quantum Overlaps}

Reference \cite{Maho15a} showed that the compression advantage $\Cmu - C_q(L)$
is determined by quantum overlaps between signal states in the \qML. Moreover,
these overlaps are determined by words whose state paths merge.

\begin{figure}
\includegraphics[width=0.6\linewidth]{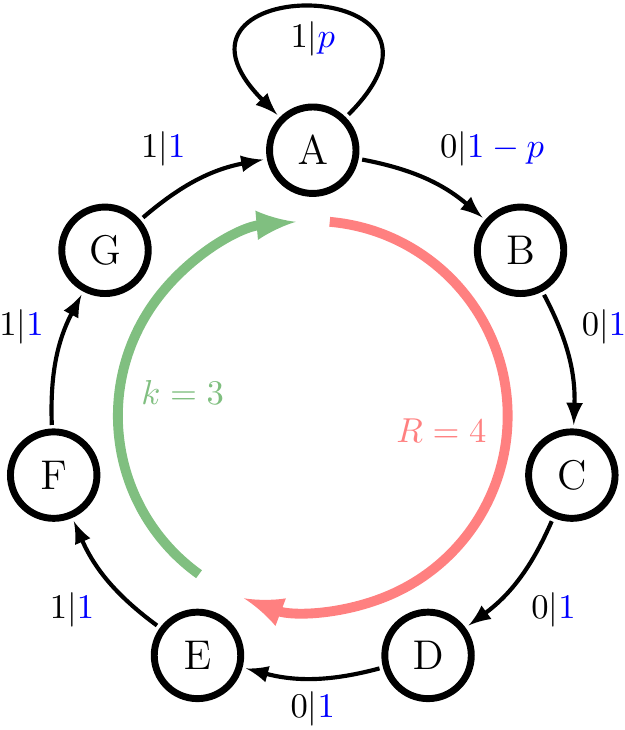}
\caption{\EM\ for the ($4$--$3$)-Golden Mean Process: The cycle's red part
  indicates the ``Markov'' portion and the green, the ``cryptic'' portion.
  The time scales $R$ and $k$ are tuned by changing the lengths of these
  two parts. Edges labeled $\ms|p$ denote taking the state-to-state transition
  with probability $p$ while emitting symbol $\ms \in \Abet$.
  }
\label{fig:43GM_eM}
\end{figure}

To illustrate, we compute some overlaps for the ($R$--$k$)-Golden Mean Process,
showing how they depend on $L$. (See Fig. \ref{fig:43GM_eM} for its \eM\ state
transition diagram.) This process was designed to have a tuneable Markov order
$R$ and cryptic order $k$; here we choose $R=4$ and $k=3$. (Refer to Ref.
\cite{Jame10a} for more on this process and detailed discussion of Markov and
cryptic orders.)

At length $L=0$, each signal state is simply the basis state corresponding to
its causal state: $\ket{\eta_i(0)} = \ket{\st_i}$. Since the \eM\ is minimal,
there are no overlaps in the state vectors.

At length $L=1$ codewords, we find the first nontrivial overlap. This
corresponds to paths $A \xrightarrow{1} A$ and $G \xrightarrow{1} A$ merging at
state $A$ and we have:
\begin{align*}
\ket{\eta_A(1)} &= \sqrt{p} \ket{1A} + \sqrt{1-p} \ket{0B} ~\text{and}\\
\ket{\eta_G(1)} &= \ket{1A}
 ~.
\end{align*}
This yields the overlap:
\begin{align*}
\braket{\eta_A(1) | \eta_G(1)} &= \sqrt{p}.
\end{align*}

Going on to length $L=2$ codewords, more overlaps arise from mergings of
more state paths. The three quantum signal states:
\begin{align*}
\ket{\eta_A(2)} &= p \ket{11A} + \sqrt{p(1-p)} \ket{10B} + \sqrt{(1-p)}
\ket{00C} ~,\\
\ket{\eta_F(2)} &= \ket{11A} ~,~\text{and}\\
\ket{\eta_G(2)} &= \sqrt{p} \ket{11A} + \sqrt{1-p} \ket{10B}
\end{align*}
interact to yield the overlaps:
\begin{align*}
\braket{\eta_A(2) | \eta_F(2)} &= p ~,\\
\braket{\eta_F(2) | \eta_G(2)} &= \sqrt{p} ~,~\text{and}\\
\braket{\eta_A(2) | \eta_G(2)} &= p\sqrt{p} + (1-p) \sqrt{p} = \sqrt{p}
  ~.
\end{align*}
The overlaps between $(A,F)$ and $(F,G)$ are new. The $(A,G)$ overlap has the
same value as that for $(F,G)$, however its computation at $L=2$ involved two
terms instead of one. This is because no new merger has occurred; the $L=1$
merger, affected by symbol $1$, was simply propagated forward along two
different state paths having prefix $1$. The redundant paths are: $A
\xrightarrow{10} B$ overlaps $G \xrightarrow{10} B$ and $A \xrightarrow{11} A$
overlaps $G \xrightarrow{11} A$. A naive computation of overlaps must contend
with this type of redundancy.

\section{Quantum Pairwise-Merger Machine}

To calculate signal-state overlaps, we introduce the quantum pairwise-merger
machine, a transient graph structure that efficiently encapsulates the
structure of state paths. As we saw in the example, computation of overlaps
amounts to tracking state-path mergers. It is important that we do this in a
systematic manner to avoid redundancies. The new machine does just this.

We begin by first constructing the \emph{pairwise-merger machine} (PMM),
previously introduced to compute overlaps \cite{Maho15a}. There, probabilities
were computed for each word found by scanning through the PMM. This method
significantly reduced the number of words from the typically exponential large
number in a process' language and also gave a stopping criterion for PMMs with
cycles. This was a vast improvement over naive constructions of the
signal-state ensemble (just illustrated) and over von Neumann entropy
calculation via diagonalization of an ever-growing state space.

Appropriately weighting PMM transitions yields the \emph{quantum
PMM} (QPMM), which then not only
captures which states merge given which words, but also the contribution each
merger makes to a quantum overlap. The QPMM has one obvious advantage over the
PMM. The \emph{particular} word that produces an overlap is ultimately
unimportant; only the amount of overlap generated is important.
Therefore, summing over symbols in the QPMM to obtain its internal state
transitions removes this combinatorial factor. Appreciating other
significant advantages to this matrix-based approach requires more development first.

To build the QPMM from a given process' \eM:
\begin{enumerate}
\setlength{\topsep}{0pt}
\setlength{\itemsep}{0pt}
\setlength{\parsep}{0pt}
\item Construct the set of (unordered) pairs of (distinct) \eM\ states: 
	$(\st_i, \st_j)$. We call these ``pair-states''. To this set, add a special
	state called SINK (short for ``sink of synchronization'') which is the
	terminal state.
\item For each pair-state $(\st_i, \st_j)$ and each symbol $x \in \Abet$, there
	are three cases to address:
\begin{enumerate}
\setlength{\topsep}{0pt}
\setlength{\itemsep}{0pt}
\setlength{\parsep}{0pt}
\item If at least one of the two \eM\ states $\st_i$ or $\st_j$ has no outgoing 
	transition on symbol $x$, then do nothing.
\item If both \eM\ states $\st_i$ and $\st_j$ have a transition on symbol $x$
	to the same state $\st_m$, then connect pair-state $(\st_i, \st_j)$ to SINK
	with an edge labeled $x$. This represents a merger. 
\item If both \eM\ states $\st_i$ and $\st_j$ have a transition on symbol $x$
	to two distinct \eM\ states $\st_m$ and $\st_n$ where $m \neq n$, then
	connect pair-state $(\st_i, \st_j)$ to pair-state $(\st_m, \st_n)$ with an
	edge labeled $x$. (There are no further restrictions on $m$ and $n$.)
\end{enumerate}
\item Remove all edges that are not part of a path that leads to SINK.
\item Remove all pair-states that do not have a path to SINK.
\end{enumerate}
This is the PMM. Now, add information about transition probabilities to this topological structure to obtain the QPMM:
\begin{enumerate}
\setcounter{enumi}{4}
\item For each pair-state $(\st_i, \st_j)$ in the PMM, add to each 
	outgoing edge the weight $\sqrt{\Pr(x | \st_i) \Pr(x | \st_j)}$, where $x$
	is the symbol associated with that edge. Note that two states in QPMM may
	be connected with multiple edges (for different symbols).
\end{enumerate}

\begin{figure}
\includegraphics[width=0.6\linewidth]{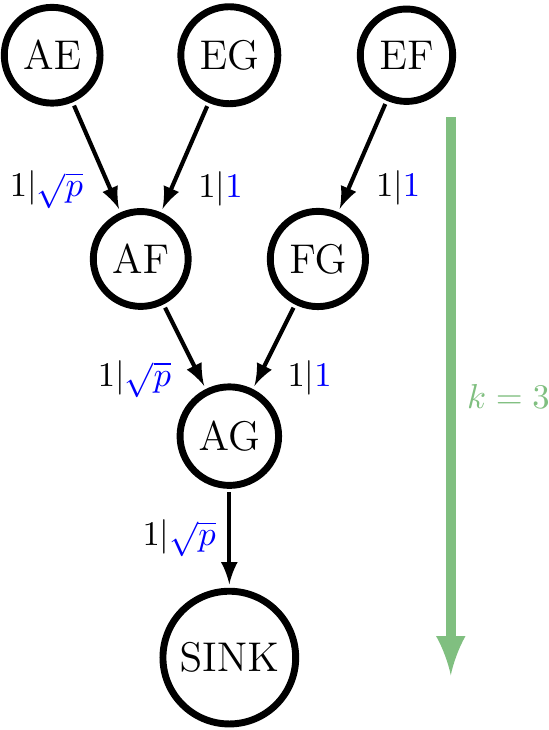}
\caption{QPMM for the ($4$--$3$)-Golden Mean Process. Its depth is related to
	the cryptic order $k$.
 }
\label{fig:43GM_QPMM}
\end{figure}

Returning to our example, Fig.~\ref{fig:43GM_QPMM} gives the QPMM for the
(4--3)-Golden Mean Process. Using it, we can easily determine the length at
which a contribution is made to a given overlap. We consider codeword lengths
$L = 1, 2, \ldots$ by walking up the QPMM from SINK. For example, pair $(A,G)$
receives a contribution of $\sqrt{p}$ at $L=1$. Furthermore, $(A,G)$ receives
no additional contributions at larger $L$. Pairs $(A,F)$ and $(F,G)$, though,
receive contributions $p = \sqrt{p} \times \sqrt{p}$ and $\sqrt{p} = \sqrt{p}
\times 1$ at $L=2$, respectively.

The QPMM is \emph{not} a HMM, since the edge weights do not yield a stochastic
matrix. However, like a HMM, we can consider its labeled ``transition''
matrices $\{ \zeta^{(\ms)} \}$, $\ms \in \Abet$. Just as for their classical
\eM\ counterparts, we index these matrices such that $\zeta^{(\ms)}_{u,v}$
indicates the edge going from pair-state $u$ to pair-state $v$. Since the
overlap contribution, and not the inducing word, is of interest, the important
object is simply the resulting state-to-state substochastic matrix $\zeta =
\sum_{\ms \in \Abet} \zeta^{(\ms)}$. The matrix $\zeta$ is the heart of our
closed-form expressions for quantum coding costs, which follow shortly. As we
noted above, it is this step that greatly reduces the combinatorial growth of
paths that would otherwise make the calculations unwieldy.

To be explicit, our (4--3)-Golden Mean Process has:
\begin{align*}
\zeta = 
\bordermatrix{
\phantom{XX} & AE & EG & EF & AF & FG & AG & \text{SINK} \cr
AE & 0 & 0 & 0 & \sqrt{p} & 0 & 0 & 0 \cr
EG & 0 & 0 & 0 & 1 & 0 & 0 & 0 \cr
EF & 0 & 0 & 0 & 0 & 1 & 0 & 0 \cr
AF & 0 & 0 & 0 & 0 & 0 & \sqrt{p} & 0 \cr
FG & 0 & 0 & 0 & 0 & 0 & 1 & 0 \cr
AG & 0 & 0 & 0 & 0 & 0 & 0 & \sqrt{p} \cr
\text{SINK} & 0 & 0 & 0 & 0 & 0 & 0 & 0\cr
} ~.
\end{align*} 

\section{Overlaps from the QPMM}

As we saw in the example, overlaps accumulate contributions as ``probability amplitude'' is pushed through the QPMM down to the SINK. Each successive overlap augmentation can thus be expressed in terms of the next iterate of $\zeta$:
\begin{align*}
\braket{\eta_i(L) | \eta_j(L)} & - \braket{\eta_i{(L-1)} | \eta_j(L-1)} \\
   & \quad\quad = \braket{(\st_i, \st_j) | \zeta^L | \synk}
   ~.
\end{align*}
The general expression for quantum overlaps follows immediately:
\begin{align}
\braket{\eta_i(L) | \eta_j(L)} = \braket{(\st_i, \st_j) | \sum \limits_{n=0}^L \zeta^n | \synk}
  ~.
\label{eq:OverlapsGeneral}
\end{align}
This form makes clear the cumulative nature of quantum overlaps and the fact
that overlap contributions are not ``labeled''.

Note that there are two trivial overlap types. Self-overlaps are always 1; this
follows from Eq.~\eqref{eq:OverlapsGeneral} since $\bra{(\st_i, \st_i)} = \bra{ \synk }$. Overlaps with no corresponding
pair-state in the QPMM are defined to be zero for all $L$.


Now, we show that there are two behaviors that contribute to overlaps: a
finite-horizon component and an infinite-horizon component. Some processes have
only one type or the other, while many have both. We start with the familiar
($R$--$k$)-GM, which has only the finite-horizon behavior.

\subsection{Finite Horizon: ($R$--$k$)-Golden Mean Process}

Let's use the general expression Eq.~\eqref{eq:OverlapsGeneral} to compute the
Hermitian positive semidefinite \emph{overlap matrices} $A_L A_L^\dagger$
with components $(A_L A_L^\dagger )_{i, j} = \braket{\eta_i(L) | \eta_j(L)}$
for lengths $L=1,2,3,4$ for
the ($R$--$k$)-Golden Mean Process. We highlight in blue the matrix elements
that have changed from the previous length. All overlaps begin with the
identity matrix, here $I_{7}$ as we have seven states in the \eM\
(Fig.~\ref{fig:43GM_eM}). Then, at $L=1$ we have one overlap.
The overlap matrix, with elements $\braket{\eta_i(1) | \eta_j(1)}$, is:
\begin{align*}
A_1 A_1^\dagger =
\bordermatrix{
\phantom{X} & A & B & C & D & E & F & G \cr
A & 1 & 0 & 0 & 0 & 0 & 0 & \textcolor{blue}{\sqrt{p}} \cr
B & 0 & 1 & 0 & 0 & 0 & 0 & 0 \cr
C & 0 & 0 & 1 & 0 & 0 & 0 & 0 \cr
D & 0 & 0 & 0 & 1 & 0 & 0 & 0 \cr
E & 0 & 0 & 0 & 0 & 1 & 0 & 0 \cr
F & 0 & 0 & 0 & 0 & 0 & 1 & 0 \cr
G & \textcolor{blue}{\sqrt{p}} & 0 & 0 & 0 & 0 & 0 & 1\cr
}
  ~.
\end{align*} 
Next, for $L=2$ we find two new overlaps.  
The overlap matrix, with elements $\braket{\eta_i(2) | \eta_j(2)}$, is:
\begin{align*}
A_2 A_2^\dagger =
\bordermatrix{
\phantom{X} & A & B & C & D & E & F & G \cr
A & 1 & 0 & 0 & 0 & 0 & \textcolor{blue}{p} & \sqrt{p} \cr
B & 0 & 1 & 0 & 0 & 0 & 0 & 0 \cr
C & 0 & 0 & 1 & 0 & 0 & 0 & 0 \cr
D & 0 & 0 & 0 & 1 & 0 & 0 & 0 \cr
E & 0 & 0 & 0 & 0 & 1 & 0 & 0 \cr
F & \textcolor{blue}{p} & 0 & 0 & 0 & 0 & 1 & \textcolor{blue}{\sqrt{p}} \cr
G & \sqrt{p} & 0 & 0 & 0 & 0 & \textcolor{blue}{\sqrt{p}} & 1 \cr
}
  ~.
\end{align*} 
For $L=3$, there are three new overlaps.
The overlap matrix, with elements $\braket{\eta_i(3) | \eta_j(3)}$, is:
\begin{align*}
A_3 A_3^\dagger =
\bordermatrix{
\phantom{X} & A & B & C & D & E & F & G \cr
A & 1 & 0 & 0 & 0 & \textcolor{blue}{\sqrt{p}^3} & p & \sqrt{p} \cr
B & 0 & 1 & 0 & 0 & 0 & 0 & 0 \cr
C & 0 & 0 & 1 & 0 & 0 & 0 & 0 \cr
D & 0 & 0 & 0 & 1 & 0 & 0 & 0 \cr
E & \textcolor{blue}{\sqrt{p}^3} & 0 & 0 & 0 & 1 & \textcolor{blue}{\sqrt{p}} & \textcolor{blue}{p} \cr
F & p & 0 & 0 & 0 & \textcolor{blue}{\sqrt{p}} & 1 & \sqrt{p} \cr
G & \sqrt{p} & 0 & 0 & 0 & \textcolor{blue}{p} & \sqrt{p} & 1 \cr
}
  ~.
\end{align*} 
Finally, for $L=4$, we find the same matrix as $L=3$: $\braket{\eta_i(4) |
\eta_j(4)} = \braket{\eta_i(3) | \eta_j(3)}$ for all $i$ and $j$. And, in fact,
this is true for all $L \geq 3$. Therefore, all overlap information has been
uncovered at codeword length $L = 3$.

Looking at the QPMM in Fig.~\ref{fig:43GM_QPMM}, we recognize that the
saturation of the overlap matrix corresponds to the finite \emph{depth} $d$ of
the directed graph---the longest state-path through the QPMM that ends in the
SINK state. Equivalently, the depth corresponds to the nilpotency of $\zeta$:
\begin{align}
d = \min \bigl\{ n \in \mathbb{N} : \zeta^{n} = 0 \bigr\} ~.
\end{align}
Note that the ($4-3$)-Golden Mean Process QPMM is a tree of depth $4$.

Whenever the QPMM is a tree or, more generally, a directed-acyclic graph (DAG),
the overlaps will similarly have a finite-length horizon equal to the depth $d$.
The nilpotency of $\zeta$ for finite-depth DAGs allows for a truncated form of
the general overlap expression Eq.~(\ref{eq:OverlapsGeneral}):
\begin{align}
\braket{\eta_i(L) | \eta_j(L)} = \braket{(\st_i, \st_j) | \! \sum \limits_{n=0}^{\min(L, d -1 )} \! \!\zeta^n \, \, | \synk}
  ~.
\label{eq:TruncatedOverlap_wDepth}
\end{align}
This form is clearly advantageous for any process whose QPMM is a finite DAG.
Naturally then, we are led to ask: What property of a process leads to a finite
DAG? To answer this question, we reconsider how overlap is accumulated via the
merging of state-paths.

\begin{figure}
\includegraphics[width=\linewidth]{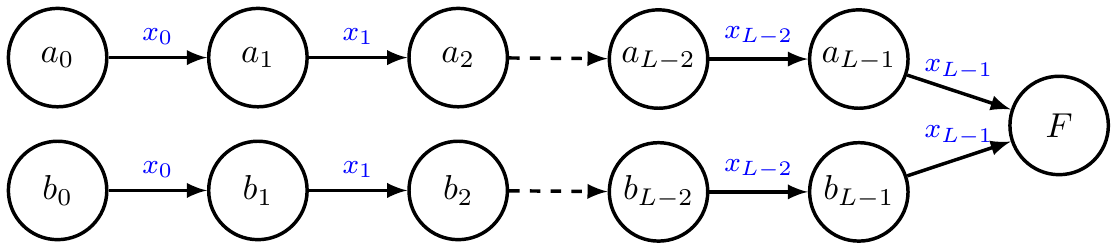}
\caption{$L$-merge: Two causal-state paths---$(a_0, \ldots, a_{L-1}, F)$ and
  $(b_0, \ldots, b_{L-1}, F)$ where states $a_i \neq b_i$, for all $i \in [0,
  \, L-1]$---generate the same word $w = \ms_0 \ms_1 \ldots \ms_{L-1}$ and
  merge only on the last output symbol $\ms_{L-1}$ into a common final state
  $F$.
  }
\label{fig:Lmerge}
\end{figure}

Paths through the QPMM represent causal-state-path mergers. To make this more
precise, we introduce the concept of an $L$-merge, which is most intuitively
understood through Fig.~\ref{fig:Lmerge}:
\begin{Def}
An \emph{$L$-merge} consists of a length-$L$ word $w$ and two state paths each
of length $L+1$ that each generate the word $w$ and end at the same state
exactly at length $L+1$. We denote the word $w = (x_0, \ldots, x_{L-1})$ and
state paths $(a_0, \ldots, a_{L-1}, F)$ and $(b_0, \ldots, b_{L-1}, F)$ where
states $a_i \neq b_i$, for all $i \in [0, \, L-1]$ and, trivially, $F=F$, the
final state in which the paths end.
\end{Def}

Immediately, we see that every labeled path of length-$L$ through the QPMM that ends in SINK is an $L$-merge.

Such causal-state-path merging not only contributes to quantum overlap, but
also contributes to a process' crypticity $\H[\St_0 | \MS_{0:\infty}]$, which is
accumulated at all lengths up to the \emph{cryptic order} \cite{Crut10a}:
\begin{Def}
The \emph{cryptic order} $k$ of a process is the minimum length $L$ for which
$\H[\St_L | \MS_{0:\infty}] = 0$.
\end{Def}

A process' \emph{Markov order} is:
\begin{align*}
R = \min \{L : \H[\St_L | \MS_{0:L}] = 0\}
  ~.
\end{align*}
A more familiar length-scale characterizing historical dependence, $R$ depends
on both path merging and path termination due to disallowed transitions. The
cryptic order, in contrast, effectively ignores the termination events and is
therefore upper-bounded by the Markov order: $k \leq R$. This bound is also
easy to see given the extra conditional variable $\MS_{L:\infty}$ in crypticity
($\MS_{0:\infty} = \MS_{0:L} \MS_{L:\infty}$) \cite{Maho09a, Maho11a}.

The following lemma states a helpful relation between cryptic order and $L$-merges.
{\Lem \label{lem:Lmerge} Given an \eM\ with cryptic order $k$: for $L \leq k$, there exists an $L-$merge; 
for $ L > k$, there exists no $L-$merge.}

{\ProLem See App.~\ref{APPOVER}.}

Each $L$-merge corresponds with a real, positive-semidefinite contribution to
some quantum overlap. By Lemma~\ref{lem:Lmerge}, for a cryptic-order $k$
process there is at least one $L$-merge at each length $L \in [1, \ldots, k]$
and none beyond $k$. Therefore, at least one overlap receives a real,
positive-semidefinite contribution at each length up until $k$, where there are
no further contributions. This leads to our result for overlap accumulation and
saturation in terms of the cryptic order.

{\The \label{thm:qq} 
Given a process with cryptic order $k$, for each $L \in [0,k]$, each quantum
overlap is a nondecreasing function of $L$:
\begin{align*}
\braket{\eta_i(L+1)| \eta_j(L+1)} \geq \braket{\eta_i(L)| \eta_j(L)}
  ~.
\end{align*}
Furthermore, for each $L \in [1,k]$, there exists at least one overlap that is
increased, as a result of a corresponding $L$-merge. For all remaining $L \geq k$, each overlap takes the constant value $\braket{\eta_i(k)| \eta_j(k)} $.
}

{\ProLem See App.~\ref{APPOVER}.}

Evidently, the cryptic order is an important length scale not only for
classical processes, but also when building efficient quantum encoders.

As an important corollary, this theorem also establishes the relation between cryptic order of the process and depth of its QPMM:
\begin{align}
d = k+1 ~.
\end{align}
Thus, we discovered the process property that corresponds to a finite DAG
QPMM is finite cryptic order. Moreover, the cryptic order corresponds to a
topological feature of the QPMM, the depth $d$, responsible for saturation of
the overlaps.

This leads to rephrasing the truncated form of the overlaps sum:
\begin{align}
\braket{\eta_i(L) | \eta_j(L)} = \braket{(\st_i, \st_j) | \! \sum \limits_{n=0}^{\min(L,k)} \zeta^n \, | \synk}.
\label{eq:TruncatedOverlap}
\end{align}
This form is advantageous for any process that is finite cryptic order. This,
of course, includes all finite Markov-order processes---processes used quite
commonly in a variety of disciplines.

Since the quantum-compression advantage $C_q(L)$ is a function of only $\boldsymbol{\pi}$ and quantum overlaps, the preceding development also gives a direct lesson about the $C_q(L)$ saturation.

{\Cor \label{thm:CqLqeqk} $C_q(L)$ has constant value $C_q(k)$  for $L \geq k$.}

{\ProCor The entropy of an ensemble of pure signal states $\{ p_i ,
\ket{\psi_i} \}$ is a function of only probabilities $p_i$ and overlaps
$\{\braket{\psi_i|\psi_j}\}$. The result then follows directly from
Thm.~\ref{thm:qq}.}

Having established connections among depth, cryptic order, and saturation, we
seem to be done analyzing quantum overlap---at least for the finite-cryptic
case. To prepare for going beyond finite horizons, however, we should reflect
on the spectral origin of $\zeta$'s nilpotency.

A nilpotent matrix, such as $\zeta$ in the finite-cryptic case, has only the
eigenvalue of zero. This can perhaps most easily be seen if the pair-states are
ordered according to their distance from SINK, so that $\zeta$ is triangular
with only zeros along the diagonal.

However, for finite DAGs with depth $ d > 1$, the standard
eigenvalue--eigenvector decomposition is insufficient to form a complete
basis---the corresponding $\zeta$ is necessarily nondiagonalizable due to the
geometric multiplicity of the zero eigenvalue being less than its algebraic
multiplicity. \emph{Generalized eigenvectors} must be invoked to form a
complete basis \cite{Franklin00}. Intuitively, this type of
nondiagonalizability can be understood as the intrinsic interdependence among
pair-states in propagating probability amplitude through a branch of the DAG.
When $\zeta$ is rendered into Jordan block form via a similarity
transformation, the size of the largest Jordan block associated with the zero
eigenvalue is called the \emph{index} $\nu_0$ of the zero eigenvalue. It turns
out to be equal to the depth for finite DAGs.

Summarizing, the finite-horizon case is characterized by several related
features: (i) the QPMM is a DAG (of finite depth), (ii) the depth of the QPMM
is one greater than the cryptic order, (iii) the matrix $\zeta$ has only the
eigenvalue zero, and (iv) the depth is equal to the index of this
zero-eigenvalue, meaning that $\zeta$ has at least $k$ generalized
eigenvectors.  More generally, $\zeta$ can have other eigenvalues and this
corresponds to richer structure that we explore next.

\subsection{Infinite Horizon: Lollipop Process}

Now we ask, what happens when the QPMM is not a directed acyclic graph?
That is, what happens when it contains \emph{cycles}?

It is clear that the depth $d$ diverges, implying that the cryptic order is
infinite. Therefore, the sum in Eq.~(\ref{eq:OverlapsGeneral}) may no longer be
truncated. We also know that infinite-cryptic processes become ubiquitous as
\eM\ state size increases. Have we lost our calculational efficiencies? No, in
fact, there are greater advantages yet to be gained.

We first observe that a QPMM's $\zeta$ breaks into two pieces. One has a
finite horizon reminiscent of finite cryptic order just analyzed, and the other
has an infinite horizon, but is analytically quite tractable, as we now show.
More generally, a linear operator $A$ may be decomposed using the \emph{Dunford
decomposition}~\cite{Dunford54} (also known as the Jordan--Chevalley
decomposition) into:
\begin{align}
A = \mathcal{D} + \mathcal{N}
  ~,
\end{align}
where $\mathcal{D}$ is diagonalizable, $\mathcal{N}$ is nilpotent, and $\mathcal{D}$ and $\mathcal{N}$ commute. $\mathcal{N}$ makes the familiar
finite-horizon contribution, whereas the new $\mathcal{D}$ term has an infinite horizon:
$\mathcal{D}^n \neq 0$, for all $n < \infty$. In the context of infinite cryptic
processes, the finite horizon associated with $\mathcal{N}$ is no longer simply related
to QPMM depth nor, therefore, the cryptic order which is infinite.

The systematic way to address the new diagonalizable part is via a spectral
decomposition~\cite{Crut13a}, where the persistent leaky features of the QPMM
state probability evolution are understood as independently acting modes. It is
clear that $\zeta$ always has a nilpotent component associated with a zero
eigenvalue, due to the SINK state. Assuming that the remaining eigenspaces are
diagonalizable, the form of the overlaps becomes:
\begin{align}
\braket{\eta_i (L)| \eta_j (L)}
&= \sum_{\lambda \in \Lambda_{\zeta} \setminus \{ 0 \} } 
    \frac{1 - \lambda^{L+1} }{1 - \lambda} 
    \bra{ (\st_i, \st_j) } \zeta_\lambda  \ket{ \synk } \nonumber \\
    & \quad +
    \sum_{m=0}^{\min \{ L, \, \nu_0 - 1 \} }  \bra{ (\st_i, \st_j) } \zeta^m \zeta_0  \ket{ \synk } 
    ~,
    \label{eq:AlmostDiagSDofOverlap}
\end{align}
where $\Lambda_{\zeta}$ is the set of $\zeta$'s eigenvalues, $\zeta_{\lambda}$
are the projection operators corresponding to each eigenvalue, and $\nu_0$ is
the index of the zero eigenvalue, which is the size of its largest Jordan
block. We refer to this as the \emph{almost-diagonalizable} case. This case
covers all processes with generic parameters. Here, $\nu_0$ is still
responsible for the length of the finite-horizon component, but is no longer
directly related to QPMM depth or process cryptic order.

Note that in the finite-cryptic order case, the only projector $\zeta_0$ is
necessarily the identity. Therefore, Eq.~\eqref{eq:AlmostDiagSDofOverlap}
reduces to the previous form in Eq.~\eqref{eq:TruncatedOverlap}.

The spectral decomposition yields a new level of tractability for the
infinite-cryptic case. The infinite-horizon piece makes contributions at all
lengths, but in a regular way. This allows for direct computation of its total
contribution at any particular $L$, including $L \to \infty$.

\begin{figure}
\includegraphics[width=\linewidth]{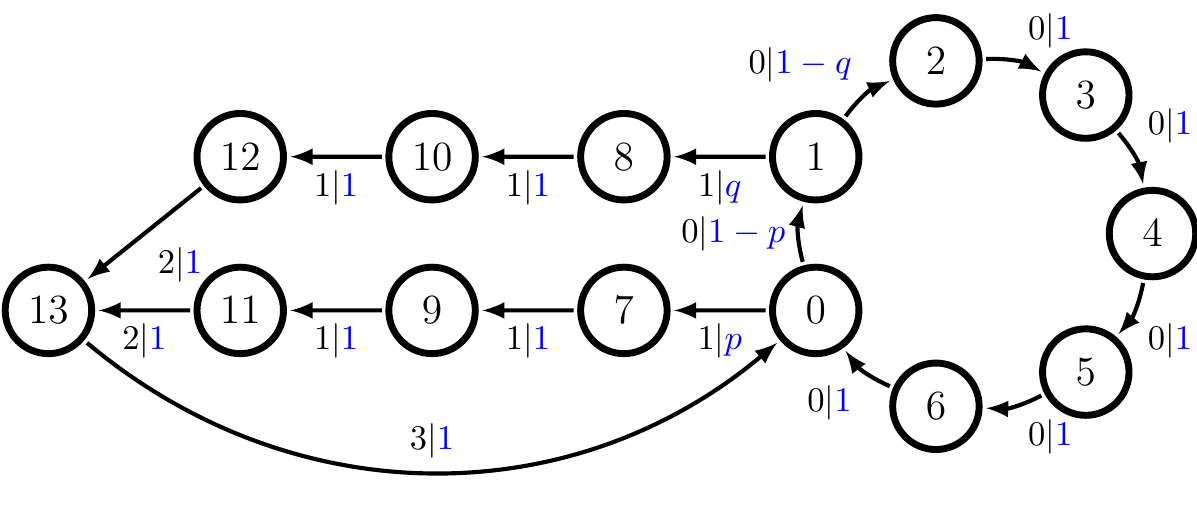}
\caption{\EM\ for the ($7$--$4$)-Lollipop Process. The cycle of $0$s leads
  to infinite Markov and cryptic orders.
  }
\label{fig:Lollipop_eM}
\end{figure}

\begin{figure}
\includegraphics[width=\linewidth]{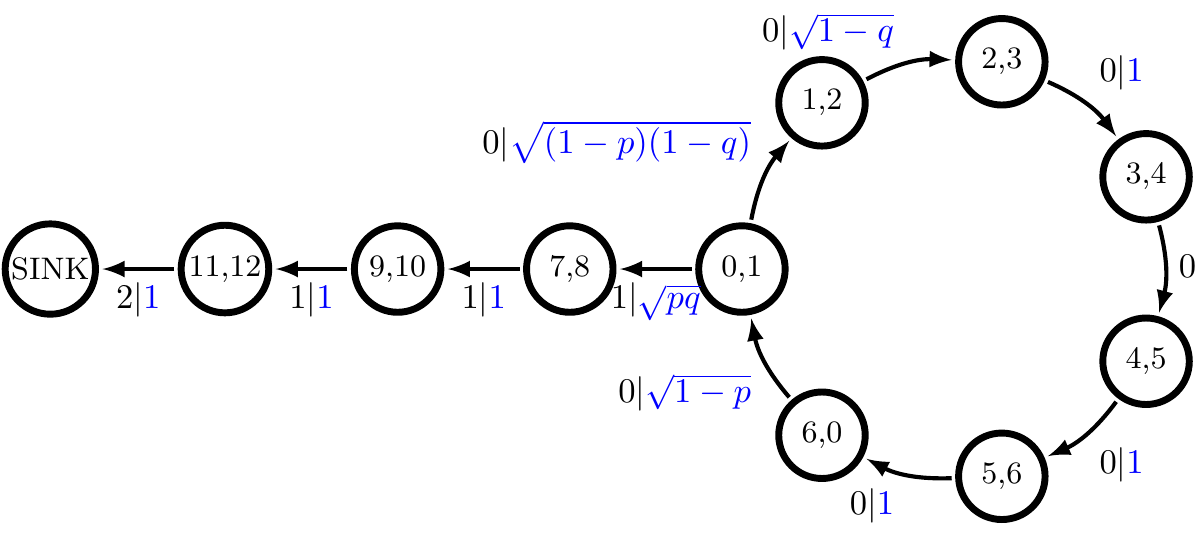}
\caption{QPMM for the (7--4)-Lollipop Process.}
\label{fig:Lollipop_QPMM}
\end{figure}

To highlight this behavior, consider the (7--4)-Lollipop Process, whose \eM\ is
shown in Fig.~\ref{fig:Lollipop_eM}. It is named for the shape of its QPMM; see
Fig.~\ref{fig:Lollipop_QPMM}. This process is a simple example of one where the
cryptic order is infinite and the finite-horizon length of the nilpotent
contribution is tunable. Roughly speaking, the diagonalizable component comes
from the ``head'' of the lollipop (the cycle), and the nilpotent part comes from the ``stick''.

It is straightforward to construct the general QPMM and thereby derive $\zeta$
for the ($N$--$M$)-Lollipop Process. Its QPMM has $N$ pair-states in a cyclic
head. The $M$ remaining pair-states constitute a finite-horizon `stick'. We
find:
\begin{align*}
\det (\zeta - \lambda I ) = (-\lambda)^{M} \bigl[ (-\lambda)^N - (1-p)(1-q) \bigr]
  ~,
\end{align*}
yielding: 
\begin{align}
\Lambda_{\zeta} &= \Bigl\{ 0 , \, \bigl[ (1-p)(1-q) \bigr]^{1/N} e^{i n 2 \pi / N }\Bigr\}_{n=0}^{N-1}
  ~,
\end{align}
with $\nu_0 = M$. 

For concreteness, consider the (7--4)-Lollipop Process with $p=q=1/2$. It has
eigenvalues $\Lambda_\zeta = \{ 0, a e^{i n \theta} \}$ and $\nu_0 = 4$, where
$a = (1/4)^{1/7}$, $\theta = 2 \pi / 7$, and $n \in \{ 0,1,2,3,4,5,6 \}$.

Each $\lambda = a e^{i n \theta}$ eigenvalue has algebraic multiplicity 1 
and associated left eigenvector:
\begin{align*}
[2 \sqrt{2} \lambda^6, \sqrt{2} \lambda^5, \lambda^4, \lambda^3, \lambda^2, \lambda^1, \lambda^0, \sqrt{2} \lambda^5, \sqrt{2} \lambda^4, \sqrt{2} \lambda^3, \sqrt{2} \lambda^2]
\end{align*}
and right eigenvector:
\begin{align*}
[ \frac{1}{2\lambda}, 1, \sqrt{2} \lambda, \sqrt{2} \lambda^2, \sqrt{2} \lambda^3, \sqrt{2} \lambda^4, \sqrt{2} \lambda^5, 0, 0, 0,0]^\top
  ~.
\end{align*}

Notice that, since $\zeta$ is not Hermitian, the right eigenvectors are not
simply the conjugate transpose of their left counterparts. The left and right
eigenvectors are fundamentally different, with the differences expressing the
QPMM's directed causal architecture.

Since each of these eigenvalues has algebraic multiplicity $1$, the
corresponding projection operators are defined in terms of right and left
eigenvectors:
\begin{align*}
\zeta_\lambda
  = \frac{\ket{\lambda} \bra{\lambda}}{\braket{\lambda | \lambda}}
  ~.
\end{align*}

The zero eigenvalue has algebraic multiplicity $\nu_0 = 4$ and geometric
multiplicity $1$, meaning that while there is only one eigenvector there are
three generalized eigenvectors. The left and right eigenvectors are:
\begin{align*}
&[0, 0, 0, 0, 0, 0, 0, 0, 0, 0, 1] ~\text{and} \\
&[0, 1, 0, 0, 0, 0, 0, -1, 0, 0, 0]^\top ~.
\end{align*}
The three generalized left eigenvectors are:
\begin{align*}
&[0, 0, 0, 0, 0, 0, 0, 1, 0, 0, 0] ~, \\
&[0, 0, 0, 0, 0, 0, 0, 0, 1, 0, 0] ~, ~\text{and}\\
&[0, 0, 0, 0, 0, 0, 0, 0, 0, 1, 0] ~; 
\end{align*}
and the three generalized right eigenvectors are:
\begin{align*}
&[0, 0,2, 0, 0, 0, 0, 0, -1, 0, 0]^\top ~, \\
&[0, 0, 0, 2 , 0, 0, 0, 0, 0,  -1, 0]^\top ~, ~\text{and}\\
&[0, 0, 0, 0, 2, 0, 0, 0, 0, 0,  -1]^\top ~.
\end{align*}

Since the index of the zero eigenvalue is larger than $1$ ($\nu_0 = 4$), the
projection operator $\zeta_0$ for the zero eigenvalue includes the
contributions from both its standard and generalized eigenvectors:
\begin{align}
\zeta_0 = \sum_{n=0}^3
  \frac{\ket{\lambda_n} \bra{\lambda_n}}{\braket{\lambda_n | \lambda_n}}
  ~,
\end{align}
where $\ket{\lambda_0}$ is the standard eigenvector and $\ket{\lambda_n}$ is
the $n^\text{th}$ generalized eigenvector for $n \geq 1$. More generally, this
sum goes over all standard and all generalized eigenvectors of the zero
eigenvalue.

Since all projection operators must sum to the identity, the projection
operator for the zero eigenvalue can be obtained alternatively from:
\begin{align}
\zeta_0 = I - \sum_{\lambda \in \Lambda_\zeta \setminus 0} \zeta_\lambda
  ~,
\end{align}
which is often a useful observation during calculations.

Using these projectors, we apply Eq.~\eqref{eq:AlmostDiagSDofOverlap} to
compute the overlaps at any $L$ for the (4--7)-Lollipop Process. This very
efficient procedure allows us to easily probe the form of quantum advantage for
any process described by a finite \eM.

Finally, we jump directly to the asymptotic overlap using the following
expression:
\begin{align}
\braket{\eta_i (\infty)| \eta_j (\infty)}
&= \bra{ (\st_i, \st_j) } \left( \sum_{n=0}^\infty \zeta^n \right) \ket{ \synk } \nonumber \\
&= \bra{ (\st_i, \st_j) } \left( I - \zeta \right)^{-1} \ket{ \synk } ~.
\label{eq:AsymptoticOverlap}
\end{align}
Note that $I - \zeta$ is invertible, since $\zeta$ is substochastic. Hence, its
spectral radius is less than unity, $1 \notin \Lambda_\zeta$, and so
$\det (I - 1 \zeta) \neq 0$. Moreover, $(I-\zeta)^{-1}$ is \emph{equal} to the
convergent Neumann series $\sum_{n=0}^\infty \zeta^n$ by Thm. 3 of Ref.
\cite[Ch.~VIII~\S~2]{Yosida95}.

Yielding an important calculational efficiency, the form of
Eq.~\eqref{eq:AsymptoticOverlap} does not require spectrally decomposing
$\zeta$ and so immediately provides the asymptotic advantage of quantum
compression. Finally, this form does not depend on the previous assumption of
$\zeta$ being almost-diagonalizable.

\section{Quantum Communication Cost}

The preceding development focused on computing overlaps between quantum signal
states for \qML\ representations of a given process. Let's not forget that the
original goal was to compute the von Neumann entropy of this ensemble---the
quantum communication cost $C_q(L)$.

The naive approach to calculating $C_q(L)$ constructs the signal states
directly and so does not make use of overlap computation. This involves working
with a Hilbert space of increasing dimension, exponential in codeword length
$L$. This quickly becomes intractable, for all but the simplest processes.  

The second approach, introduced in Ref.~\cite{Maho15a}, made use of the PMM to
compute overlaps. These overlaps were then used to construct a density operator
with those same overlaps, but in a Hilbert space of fixed size
$|\CausalStateSet|$, essentially obviating the high-dimensional embedding of
the naive approach. The elements of the resulting density matrix, however, are
nonlinear functions of the overlaps. Besides the computational burden this
entails, it makes it difficult to use the overlap matrix to theoretically infer
much about the general behavior of $C_q(L)$.

Here, we present two markedly improved approaches that circumvent these
barriers. We are ultimately interested in the von Neumann entropy which depends
only on the spectrum of the density operator. It has been pointed out that the
Gram matrix of an ensemble shares the same spectrum \cite{Jozsa00a}. The
\emph{Gram matrix} for our ensemble of pure quantum signal states is:\begin{align}
G = 
\begin{bmatrix}
\sqrt{\pi_1 \pi_1} \braket{\eta_1| \eta_1} & \cdots & \sqrt{\pi_1 \pi_{|\SSet|}} \braket{\eta_1| \eta_{|\SSet|}}\\
\vdots & \ddots & \vdots \\
\sqrt{\pi_{|\SSet|} \pi_1} \braket{\eta_{|\SSet|}| \eta_1}& \cdots &  \sqrt{\pi_1 \pi_{|\SSet|}} \braket{\eta_{|\SSet|}| \eta_{|\SSet|}}
\end{bmatrix}
  ~.
\end{align}
If we define $D_{\boldsymbol{\pi}} := \text{diag}(\boldsymbol{\pi})$, then
$G = D_{\boldsymbol{\pi}}^{1/2} A A^\dagger D_{\boldsymbol{\pi}}^{1/2}$.


Given that it is only a small step from the overlap matrix $A A^\dagger$ 
to the Gram matrix $G$, we
see the usefulness of the thorough-going overlap analysis above. The spectrum
is then computed using standard methods, symbolically or numerically, for $G$.

There is yet another surrogate matrix that also shares the spectrum but is
simpler, yet again. We call this matrix $\widetilde{G}$ the \emph{abridged-Gram
matrix}:
\begin{align}
\widetilde{G} = 
\begin{bmatrix}
\pi_1 \braket{\eta_1| \eta_1} & \cdots & \pi_1\braket{\eta_1| \eta_{|\SSet|}} \\
\vdots & \ddots & \vdots \\
\pi_{|\SSet|} \braket{\eta_{|\SSet|}| \eta_1} & \cdots &  \pi_{|\SSet|} \braket{\eta_{|\SSet|}| \eta_{|\SSet|}} 
\end{bmatrix}  ~.
\end{align}
Note that $\widetilde{G} = D_{\boldsymbol{\pi}} A A^\dagger$.

Since the spectra are identical, we can calculate $C_q(L)$ directly from the
density matrix $\rho(L)$, Gram matrix $G$, or abridged-Gram matrix
$\widetilde{G}$:
\begin{align*}
C_q(L) 
&= \, - \!  \sum_{\lambda \in \Lambda_{\rho{(L)} } } \lambda \log \lambda \\
&= \, - \!  \sum_{\lambda \in \Lambda_{G^{(L)} } } \lambda \log \lambda \\
&= \, - \!  \sum_{\lambda \in \Lambda_{\widetilde{G}^{(L)} } } \lambda \log \lambda ~.
\end{align*}
For further discussion, see App.~\ref{DMGAG}.

\begin{figure}
\includegraphics[width=\linewidth]{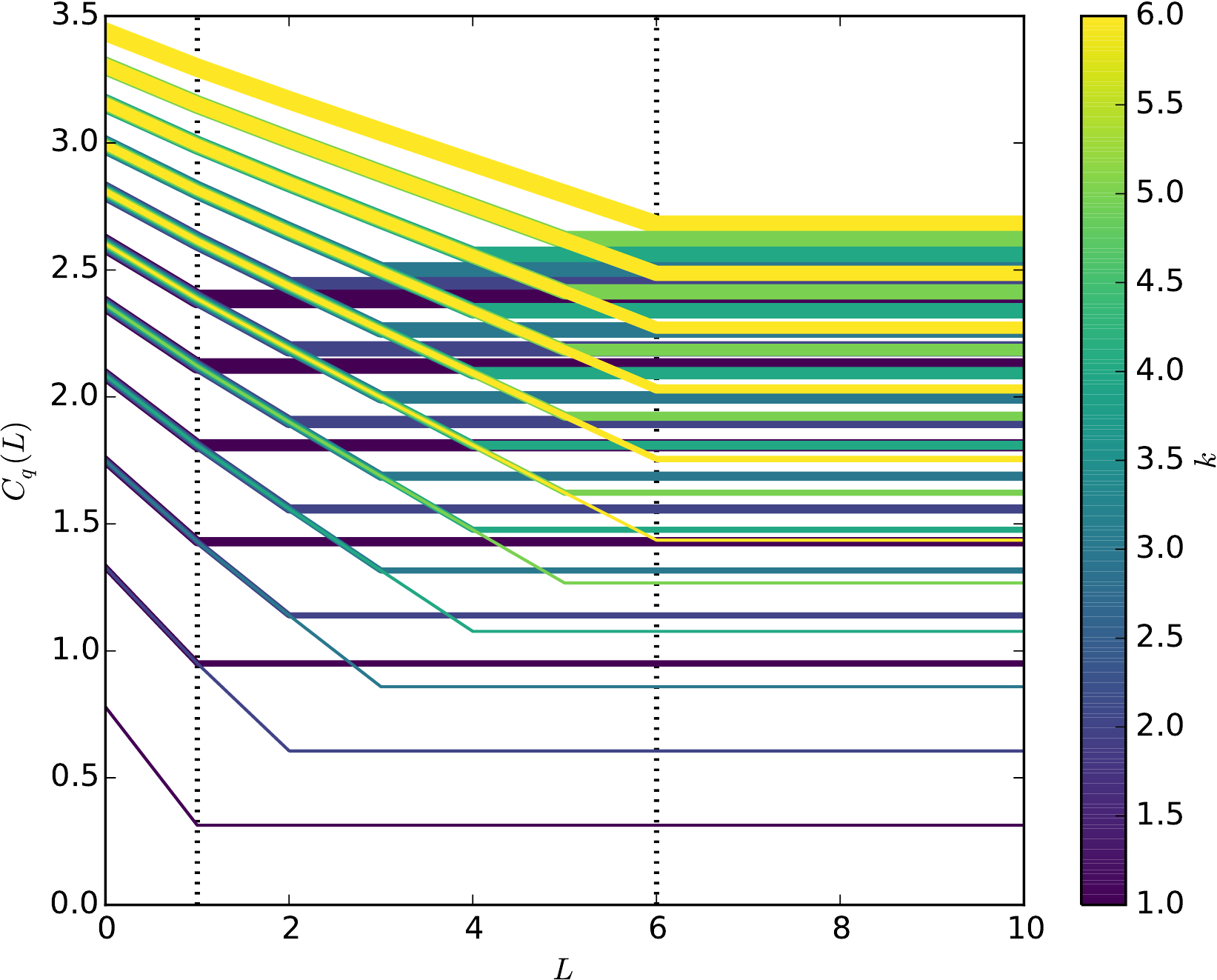}
\caption{Quantum costs $C_q(L)$ for the ($R$--$k$)-Golden Mean Process
  with $R \in \{1, \ldots, 6\}$ and $k \in \{1, \ldots, R\}$. $R$ and $k$ are
  indicated with line width and color, respectively. The probability of the
  self-loop is $p=0.7$. $C_q(L)$ roughly linearly decreases until $L=k$
  where it is then constant. Note that ($R$--$k$) agrees with
  ($(R+1)$--$(k-1)$) for $L \leq k$, as explained in App.~\ref{appex}.
  }
\label{fig:RkGM_Cq}
\end{figure}

\begin{figure}
\includegraphics[width=\linewidth]{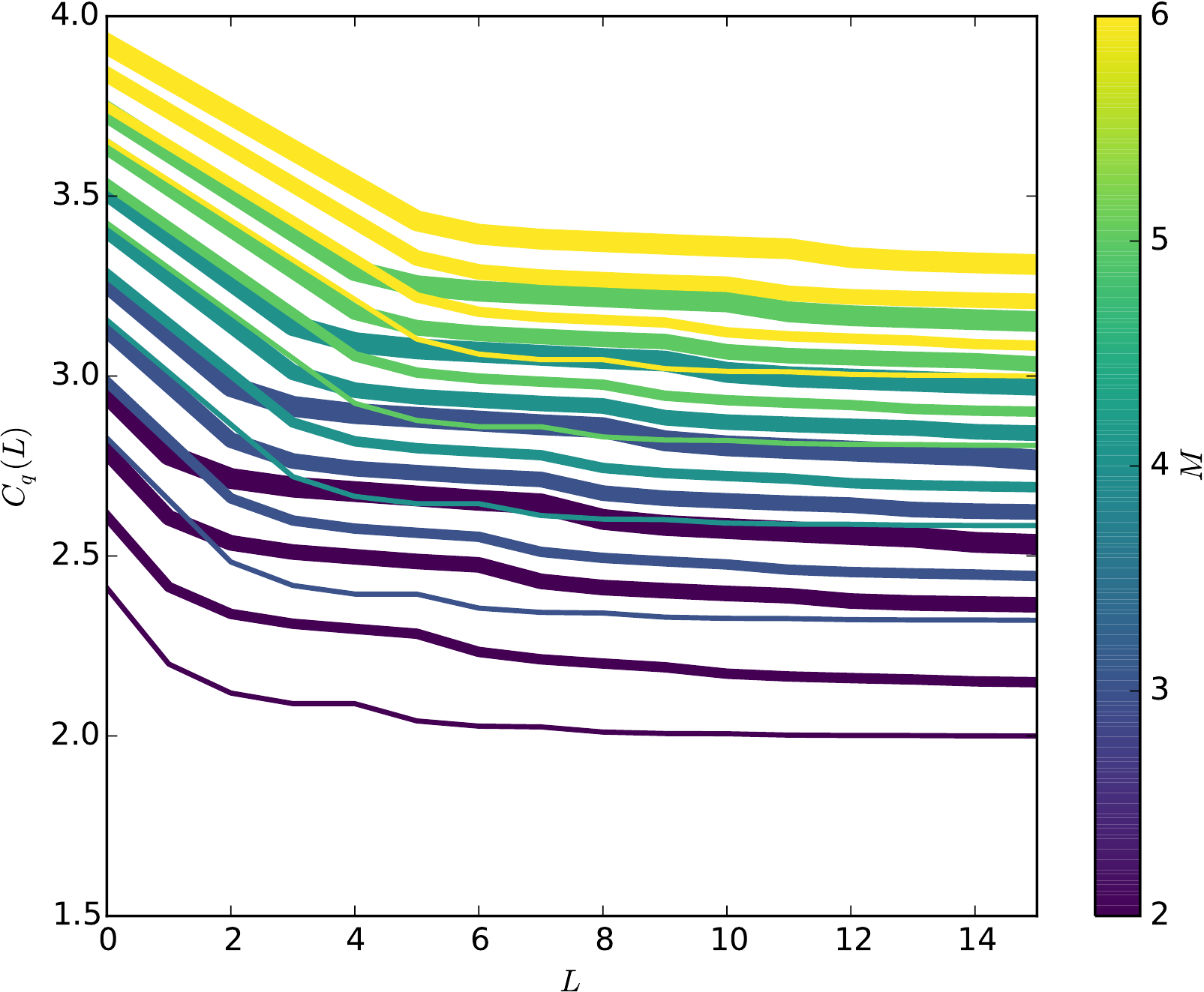}
\caption{Quantum costs $C_q(L)$ for the Lollipop process for
  $N \in \{ 3,4,5,6 \}$, $M \in \{ 2,3,4,5,6 \}$, and $p = q = 0.5$. $N$ and
  $M$ are indicated with line width and color, respectively. After a fast
  initial decrease, these curves approach their asymptotic values more slowly.
  }
\label{fig:Lollipop_Cq}
\end{figure}

Using the Gram matrix as described, we illustrate the behavior of $C_q(L)$ for
the ($R$--$k$)-Golden Mean (Fig.~\ref{fig:RkGM_Cq}) and ($N$--$M$)-Lollipop
(Fig.~\ref{fig:Lollipop_Cq}). For each of the two process classes, we compute
several instances by varying $R$ and $k$ and by varying $N$ and $M$ while
holding fixed their transition parameters. Comparing the two
figures, we qualitatively confirm the difference between a process with only a
finite-horizon contribution and one with an infinite-horizon contribution. The
($R$--$k$)-Golden Mean reaches its encoding saturation at $L=k$ the cryptic
order. The ($N$--$M$)-Lollipop only approaches this limit asymptotically.

To supplement the details already given, annotated analytic derivations of
several example processes are given in App.~\ref{appex}. These examples
serve as a pedagogical resource, with comparison and discussion of various
analytical techniques.

\section{Costs using Long Finite Codewords}

We discussed quantum-state overlaps extensively. We found that the behavior of the overlaps with $L$ is completely described through $\zeta$'s spectral decomposition. And, we showed that, for any $L$, the von Neumann entropy $C_q(L)$ can be found from the eigenvalues of the Gram matrix---a direct transformation of the overlap matrix. This is all well and good, and key progress. But, can we use this machinery to directly analyze the behavior of $C_q(L)$ as a function of $L$? For infinite-cryptic processes, the answer is an especially pleasing affirmative.

This section derives $C_q(L)$'s asymptotic behavior for large $L$; viz., $\nu_0
< L \leq k = \infty$. We show that a periodic pattern, exponentially decaying
at the rate of the largest $\zeta$-eigenvalue magnitude, dominates the
deviation from $C_q(\infty)$ for large $L$. That is, we show two things: First,
the asymptotic behavior of $C_q(L) - C_q(\infty)$ is, to first order,
exponentially decreasing as $r_1^L$, where $r_1$ is the spectral radius of
$\zeta$. Second, this exponential defines an envelope for a $\Psi$-periodic
asymptotic structure, where $\Psi$ is the least common multiple of
slowest-decaying QPMM cycle lengths.

Recall that the optimal quantum compression is given by the asymptotic von Neumann entropy:
\begin{align*}
C_q(\infty) = -\sum_{\lambda^{(\infty)} \in \Lambda_{G^{(\infty)}}} \lambda^{(\infty)} \log \bigl( \lambda^{(\infty)} \bigr) ~.
\end{align*}
We will show that when $L$ is large, $(\delta G)^{(L)} := G^{(L)} -
G^{(\infty)}$ can be treated as a perturbation to $G^{(\infty)}$.  From the
corresponding small variations $\bigl\{ (\delta \lambda)^{(L)} \bigr\}_{\lambda
\in \Lambda_G}$, direct computation of the first differential yields the
approximate change in the von Neumann entropy:
\begin{align}
(\delta S)^{(L)} = -\sum_{\lambda \in \Lambda_G}  \bigl[ \log{(\lambda^{(\infty)})} + 1 \bigr] \, ( \delta \lambda )^{(L)} ~,
\label{eq:dS}
\end{align}
so long as no zero eigenvalues of $G^{(\infty)}$ vanish at finite $L$.  Our
task, therefore, is to find $( \delta \lambda )^{(L)}$ from $(\delta G)^{(L)}$
in terms of $\zeta$'s spectral properties.

For easy reference, we first highlight our notation:
\begin{itemize}
\setlength{\topsep}{0pt}
\setlength{\itemsep}{0pt}
\setlength{\parsep}{0pt}
\item $G^{(L)}$ is a Gram matrix at length $L$ corresponding to $\rho(L)$. 
\item $\lambda^{(L)} \in \Lambda_{G^{(L)}}$ is any one of its eigenvalues.
\item $\ket{\lambda^{(L)}}$ and $\bra{\lambda^{(L)}}$ are the
	right and left eigenvectors of $G^{(L)}$ corresponding to $\lambda^{(L)}$,
	respectively.
\item $(\delta G)^{(L)} := G^{(L)} - G^{(\infty)}$ is the perturbation to
	$G^{(\infty)}$ investigated here. 
\item $\xi \in \Lambda_\zeta$ is an eigenvalue of the QPMM transition
	dynamic $\zeta$. 
\end{itemize}

If using $G$'s symmetric version, the right and left eigenvectors are
simply transposes of each other: $\bra{\lambda^{(L)}} = \bigl(
\ket{\lambda^{(L)}} \bigr)^\top$. For simplicity of the proofs, we assume
nondegeneracy of $G^{(L)}$'s eigenvalues, so that the projection operator
associated with $\lambda^{(L)}$ is $\ket{\lambda^{(L)}} \bra{\lambda^{(L)}} /
\braket{ \lambda^{(L)} | \lambda^{(L)}}$, where the denominator assures
normalization.  Nevertheless, the eigenbasis of $G^{(L)}$ is always complete,
and the final result Thm.~\ref{the:CqLAsymptoticBehavior} retains general
validity.

Here, we show that the matrix elements of $\dGL$ are arbitrarily small for large enough $L$, such that first-order perturbation is appropriate for large $L$, and give the exact form of $\dGL$ for use in the computation of
$(\delta \lambda)^{(L)}$. 

{\Prop \label{paul} For $L \geq \nu_0$, the exact change in Gram matrix is:
\begin{align*}
\dGL = - \sum_{\xi \in \Lambda_\zeta \setminus 0}
  \frac{\xi^{L+1}}{1-\xi} \, C_\xi
  ~,
\end{align*}	
where $C_\xi$ is independent of $L$ and has matrix elements:
\begin{align*}
(C_\xi)_{i, j}
  = \sqrt{\pi_i \pi_j} \braket{(\st_i, \st_j) | \zeta_\xi | \synk }
  ~.
\end{align*}	
}

{\ProProp We calculate:
\begin{align*}
(\delta G)^{(L)}_{i, j}
&= G^{(L)}_{i, j} - G^{(\infty)}_{i, j} \\
&= \sqrt{\pi_i \pi_j} \left( \braket{\eta_i^{(L)} | \eta_j^{(L)}} - \braket{\eta_i^{(\infty)} | \eta_j^{(\infty)}} \right) \\
&= - \sqrt{\pi_i \pi_j} \braket{(\st_i, \st_j) | \zeta^{L+1} (1 - \zeta)^{-1} | \synk } ~.
\end{align*}
If we assume that all nonzero eigenvalues of $\zeta$ correspond to diagonalizable subspaces, then
for $L \geq \nu_0$, the elements of $(\delta G)^{(L)}$ have the spectral decomposition:
\begin{align*}
(\delta G)^{(L)}_{i, j} = - \sum_{\xi \in \Lambda_\zeta \setminus 0} \frac{\xi^{L+1}}{1-\xi}  \sqrt{\pi_i \pi_j} \braket{(\st_i, \st_j) | \zeta_\xi | \synk } ~.
\end{align*}
Since this decomposition is common to all matrix elements, we can factor out
the $\left\{ \tfrac{\xi^{L+1}}{1-\xi} \right\}_{\xi}$, leaving the $L$-independent set of matrices: 
\begin{align*}
\Bigl\{ C_\xi : (C_\xi )_{i,j} = \sqrt{\pi_i \pi_j} \braket{(\st_i, \st_j) | \zeta_\xi | \synk }   \Bigr\}_{\xi \in \Lambda_{\zeta}}
  ~,
\end{align*}
such that:
\begin{equation*}
\dGL = - \sum_{\xi \in \Lambda_\zeta \setminus 0} \frac{\xi^{L+1}}{1-\xi} \, C_\xi ~.
\end{equation*}	
}

{\Prop \label{prop:ExactDlambdaL}
At large $L$, the first-order correction to $\lambda^{(\infty)}$ is: 
\begin{align}
(\delta \lambda)^{(L)} = 
- \sum_{\xi \in \Lambda_\zeta \setminus 0} \frac{\xi^{L+1}}{1-\xi} \, 
\frac{ \bra{\lambda^{(\infty)}} C_\xi \ket{\lambda^{(\infty)}} }{\braket{ \lambda^{(\infty)} | \lambda^{(\infty)} }} ~.
\end{align}
}

{\ProProp 
Perturbing $G^{(\infty)}$ to $G^{(\infty)}+\dGL$, the first order change in its eigenvalues is given by:
\begin{equation}
(\delta \lambda)^{(L)} = \frac{ \bra{\lambda^{(\infty)}} \dGL \ket{\lambda^{(\infty)}} }{\braket{ \lambda^{(\infty)} | \lambda^{(\infty)} }} ~,
\label{eq:1stOrderEq}
\end{equation}
which is standard first-order nondegenerate perturbation theory familiar in quantum mechanics, with the allowance for unnormalized bras and kets.
Proposition \ref{prop:ExactDlambdaL} then
follows directly from Eq.~\eqref{eq:1stOrderEq}
and Prop.~\ref{paul}.
}

{\The \label{thm:DCqL}
At large $L$, such that $\nu_0 < L \leq k = \infty$, the first-order correction
to $C_q(\infty)$ is:
\begin{align}
&C_q(L) - C_q(\infty) 
\approx (\delta S)^{(L)} \nonumber \\
& \quad = 
\sum_{\xi \in \Lambda_\zeta \setminus 0} \frac{\xi^{L+1}}{1-\xi} 
\sum_{\lambda^{(\infty)} \in \Lambda_{G^{(\infty)}}} \! \! \braket{C_\xi } \bigl[ \log(\lambda^{(\infty)}) + 1 \bigr] ~,
\end{align}
where:
\begin{align*}
\braket{C_\xi} := 
\frac{ \bra{\lambda^{(\infty)}}  C_\xi \ket{\lambda^{(\infty)}} }{\braket{ \lambda^{(\infty)} | \lambda^{(\infty)} }} ~.
\end{align*}
}

{\ProThe 
This follows directly from Eq.~\eqref{eq:dS} and
Prop.~\ref{prop:ExactDlambdaL}.}

The large-$L$ behavior of $C_q(L) - C_q(\infty)$ is a sum of decaying complex
exponentials. And, to first order, we can even calculate the coefficient of
each of these contributions.

Notice that the only $L$-dependence in Prop.~\ref{prop:ExactDlambdaL} and 
Thm.~\ref{thm:DCqL} come in the form of exponentiating eigenvalues of the QPMM
transition dynamic $\zeta$. For very large $L$, the dominant structure implied
by Prop.~\ref{prop:ExactDlambdaL} and Thm.~\ref{thm:DCqL} can be teased out by
looking at the relative contributions from $\zeta$'s first- and second-largest
magnitude sets of eigenvalues.

Let $r_1$ be the spectral radius of $\zeta$, shared by the largest eigenvalues
$\Lambda(r_1)$: $r_1 := \max_{\xi \in \Lambda_\zeta} | \xi |$. And, let
$\Lambda(r_1) := \argmax_{\xi \in \Lambda_\zeta} | \xi |$~. Then, let $r_2$ be
the second-largest magnitude of all of $\zeta$'s eigenvalues that differs from
$r_1$: $r_2 := \max_{\xi \in \Lambda_\zeta \setminus \Lambda(r_1)} | \xi | $.
And, let $\Lambda(r_2) := \argmax_{\xi \in \Lambda_\zeta \setminus
\Lambda(r_1)} | \xi | $~. Multiple eigenvalues can belong to $\Lambda(r_1)$.
Similarly, multiple eigenvalues can belong to $\Lambda(r_2)$.

Then, $0 \leq (r_2 / r_1) < 1$ if $\zeta$ has at least one nonzero eigenvalue. This is the case of interest here since we are addressing those infinite-horizon processes with $k = \infty > \nu_0$.  Hence, as $L$ becomes large, $\left( r_2 / r_1 \right)^L $ vanishes exponentially if it is not already zero. This leads to a corollary of Prop.~\ref{prop:ExactDlambdaL}.

{\Cor \label{paul2} 
For $L \geq \nu_0$, the leading deviation from $\lambda^{(\infty)}$ is:
\begin{equation*}
(\delta \lambda)^{(L)} = - r_1^{L+1} \sum_{\xi \in \Lambda(r_1)} \frac{\left( \xi / | \xi | \right)^{L+1} }{1-\xi} \braket{C_\xi} \left[ 1 + O\Bigl( \bigl( \tfrac{r_2}{r_1} \bigr)^{L} \Bigr) \right] ~.
\end{equation*}	
}

Notice that $\xi / | \xi |$ lies on the unit circle in the complex plane. Due
to their origin in cyclic graph structure, we expect each $\xi \in
\Lambda(r_1)$ to have a phase in the complex plane that is a rational fraction
of $2 \pi$. Hence, there is some $n$ for which $ \left(\xi / | \xi | \right)^n
= 1$, for all $\xi \in \Lambda(r_1)$. The minimal such $n$, call it $\Psi$,
will be of special importance:
\begin{align}
\Psi & := \min \bigl\{ n \in \mathbb{N} : \left( \xi / |\xi| \right)^n = 1
~\text{for~all}~ \xi \in \Lambda(r_1) \bigr\}
  ~.
\label{eq:Psi}
\end{align}
Since all $\xi \in \Lambda(r_1)$ originate from cycles in $\zeta$'s graph,
we have the result that $\Psi$ is equal to the least common multiple of the
cycle lengths implicated in $\Lambda(r_1)$.

For example, if all $\xi \in \Lambda(r_1)$ come from the same cycle in the
graph of $\zeta$, then $\Psi = |\Lambda(r_1)|$ and:
\begin{align*}
\Lambda(r_1) = \bigl\{ \xi_m = r_1 e^{i m 2 \pi / |\Lambda(r_1)| } \bigr\}_{m=1}^{|\Lambda(r_1)|}
  ~.
\end{align*}
That is, $\bigl\{\xi_m / | \xi_m | \bigr\}_{m=1}^{|\Lambda(r_1)|}$ are
the $|\Lambda(r_1)|^\text{th}$ roots of unity, uniformly distributed along the
unit circle. If, however, $\Lambda(r_1)$ comes from multiple cycles in 
$\zeta$'s graph, then the least common multiple of the cycle lengths should be
used in place of $|\Lambda(r_1)|$. 

Recognizing the $\Psi$-periodic structure of $\left(\xi / | \xi | \right)^n$
yields a more informative corollary of Prop.~\ref{prop:ExactDlambdaL}:
{\Cor \label{paul3} 
For $L \geq \nu_0$, the leading deviation from $\lambda^{(\infty)}$ is:
\begin{align*}
(\delta \lambda)^{(L)} = - r_1^{L+1}
  & \sum_{\xi \in \Lambda(r_1)}
  \frac{\left( \xi / | \xi | \right)^{\!\!\!\! \mod{(L+1, \, \Psi )} }}{1-\xi} \, \braket{C_\xi} \\
  & \qquad \qquad \times
  \left[ 1 + O\Bigl( \bigl( r_2 / r_1 \bigr)^{L} \Bigr) \right]
  ~.
\end{align*}	
}
Hence:
\begin{align}
(\delta \lambda)^{(L + \Psi )} \approx r_1^{ \Psi } \, (\delta \lambda)^{(L)} ~.
\end{align}

We conclude that asymptotically a pattern---of changes in the density-matrix
eigenvalues (with period $\Psi$)---decays exponentially with decay rate of
$r_1^{\Psi}$ per period.  There are immediate implications for the pattern of
asymptotic changes in $C_q(L)$ at large $L$.

\begin{figure}
\includegraphics[width=\linewidth]{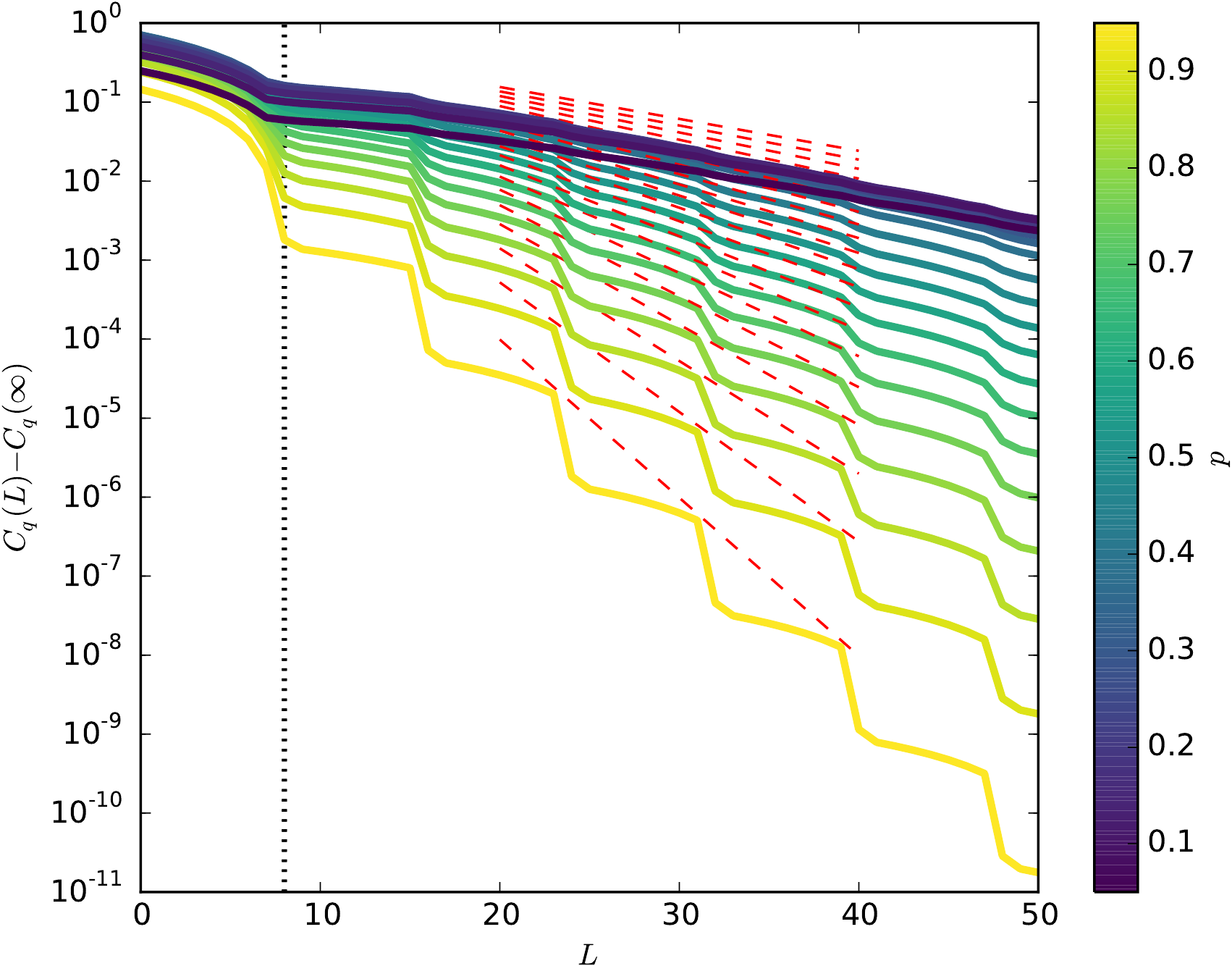}
\caption{(8,8)-Lollipop with $p \in [0.05, 0.95]$. $C_q(L) - C_q(\infty)$ on
	semilog plot illustrates asymptotically exponential behavior. Red dashed
	lines, $r_1^L$ where $r_1$ is the spectral radius of $\zeta$, quantify the
	exponential rate of decay.
	}
\label{fig:Lollipop_slope}
\end{figure}

\begin{figure}
\includegraphics[width=\linewidth]{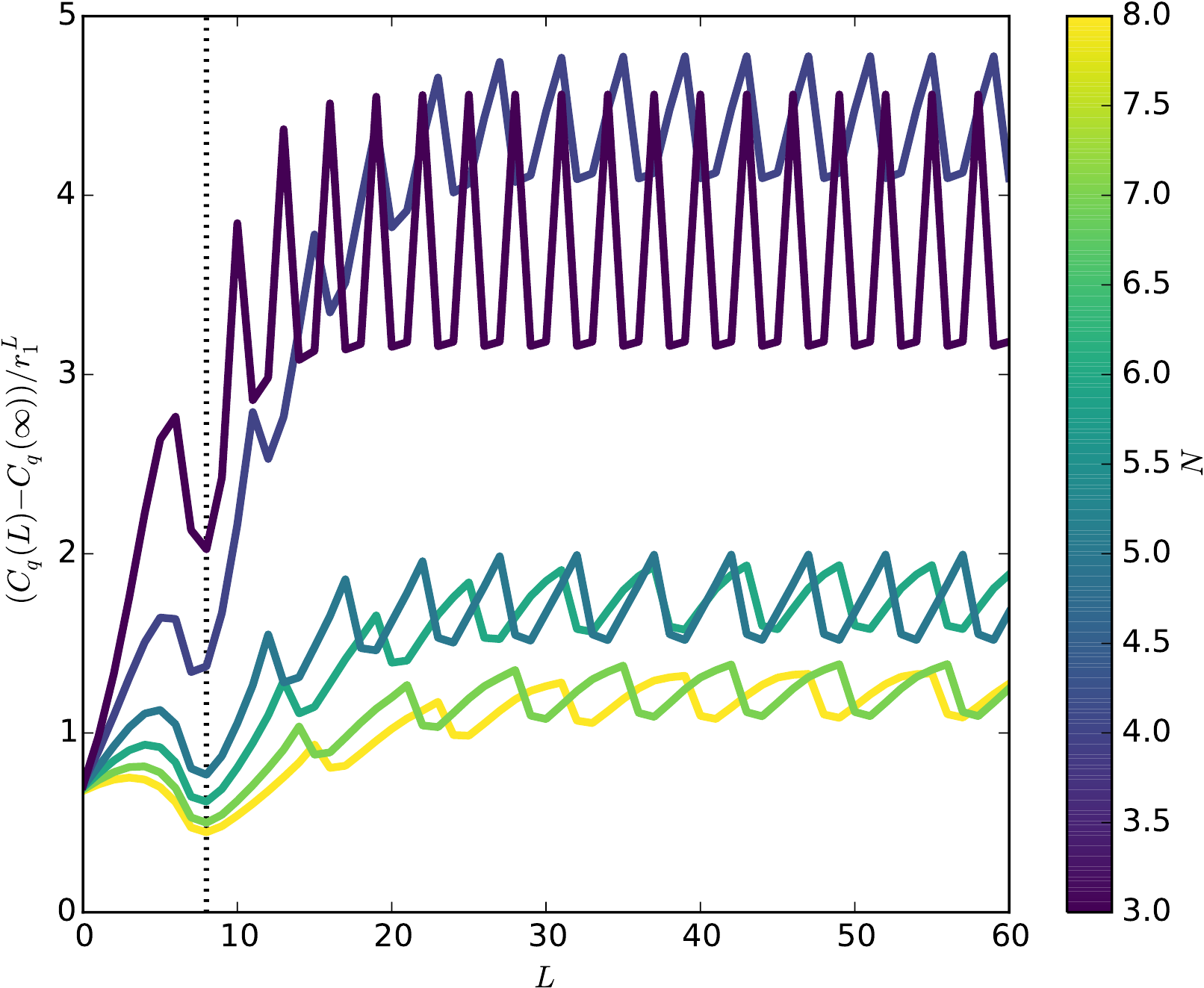}
\caption{Lollipop with $N \in \{ 3,4,5,6,7,8 \}$ and $M=8$. 
	$\bigl( C_q(L) - C_q(\infty) \bigr) / r_1^L$ demonstrates the periodicity
	of asymptotic behavior. Removing the exponential envelope makes periodicity
	of the remaining deviation more apparent.  For Lollipop, the periodicity
	$\Psi = |\Lambda(r_1)| = N$.
}
\label{fig:Lollipop_period}
\end{figure}

\begin{figure}
\includegraphics[width=\linewidth]{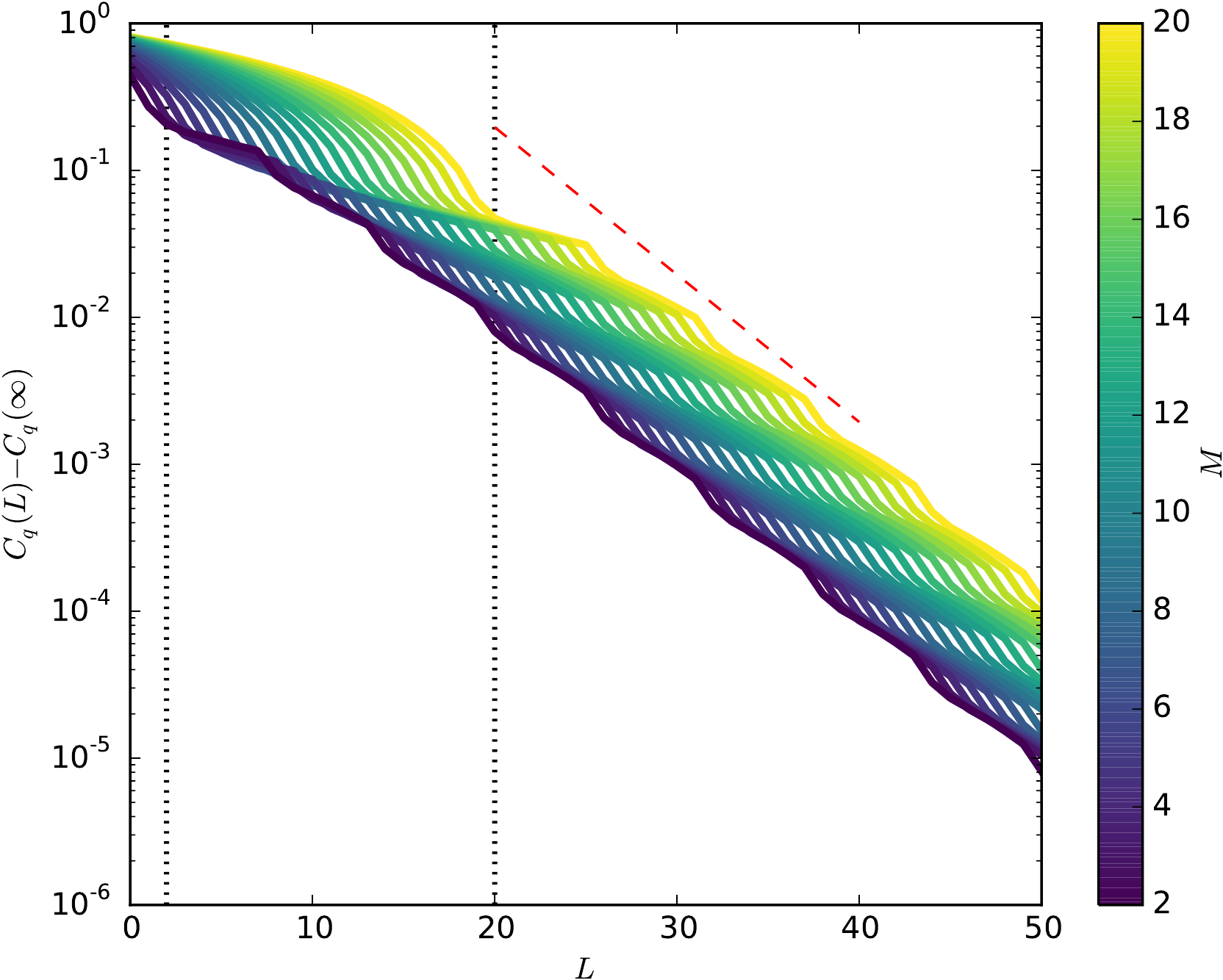}
\caption{$C_q(L) - C_q(\infty)$ for Lollipop with $N=6$ and
	$M \in \{ 2, \ldots, 20 \}$ on a semilog plot. $M$ determines the
	finite-horizon length, where the nilpotent part of $\zeta$ vanishes.
	Vertical lines indicate $L=2$ and $L=20$, the shortest and longest such
	length in this group.
	}
\label{fig:Lollipop_offset}
\end{figure}

{\Cor \label{paul4} 
For $L \geq \nu_0$, the leading deviation from $C_q(\infty)$ is:
\begin{align*}
& C_q(L) - C_q(\infty) \approx (\delta S)^{(L)} \\
  & \quad = r_1^{L+1} \sum_{\xi \in  \Lambda(r_1)}
  \frac{ \left( \xi / | \xi | \right)^{\!\!\!\! \mod{(L+1, \, \Psi)} } }{1-\xi} \\
& \qquad \qquad \times
  \sum_{\lambda^{(\infty)} \in \Lambda_{G^{(\infty)}}}
  \braket{C_\xi } \log(\lambda^{(\infty)}) 
  \left[ 1 + O\Bigl( \bigl( r_2 / r_1 \bigr)^{L} \Bigr) \right]
  ~.
\end{align*}	
}

The most profound implication of this detailed analysis can be summarized succinctly.

{\The \label{the:CqLAsymptoticBehavior}
For sufficiently large $L$:
\begin{align}
\frac{ C_q( L + \Psi ) - C_q(\infty) }{ C_q(L) - C_q(\infty) } \approx r_1^{\Psi}
  ~.
\label{eq:CqLAsymptoticBehavior}
\end{align}
}

That is, asymptotically a pattern---of changes in $C_q(L) - C_q(\infty)$ (with
period $\Psi$)---decays exponentially with decay rate of $r_1^{\Psi}$ per
period \footnote{In principle, we need to consider two cases: the pattern
decays to $C_q(\infty)$ from above or from below. In either case, the decay of
the $C_q(L) - C_q(\infty)$ pattern would be exponential. However, it is known
that $C_q(L)$ is strictly less than $\Cmu = C_q(0)$ for any $L$ for any
noncounifilar process (and equal otherwise). Hence, we expect that $C_q(L)$
always decays from above, as strongly corroborated by extensive numerical
exploration.}.

While the first-order perturbation allowed us to identify both the roles and
values of $r_1$ and $\Psi$ for any process and Coro.~\ref{paul4} would imply
Thm.~\ref{the:CqLAsymptoticBehavior}, Thm.~\ref{the:CqLAsymptoticBehavior}
actually transcends the limitations of the first-order approximation.

{\ProThe
Expanding $\log G^{(L)}$ in powers of $(G^{(L)} - I)$, then multiplying by
$-G^{(L)}$, shows that $C_q(L) = -\text{tr} \left[ G^{(L)} \log G^{(L)}
\right]$ can be written as:
\begin{align}
C_q(L) = - \sum_{n=0}^{\infty} a_n \, \text{tr} \bigl[ ( G^{(L)} )^n \bigr]
  ~, 
\label{eq:CqLTaylor}
\end{align}
for proper $a_n \in \mathbb{R}$. Using:
\begin{align}
G^{(L)} & =  \sum_{\xi \in \Lambda_{\zeta} \setminus 0}
  \frac{1 - \xi^{L+1}}{1-\xi} \, C_\xi \,
  + \!\! \sum_{m=0}^{\min \{ L, \, \nu_0 - 1 \} } C_{0,m}
  ~,
\end{align}
with appropriate constant matrices $C_{0,m}$, together with Eqs.~\eqref{eq:Psi}
and \eqref{eq:CqLTaylor} yields Thm.~\ref{the:CqLAsymptoticBehavior} with
general validity.
}

In the simplest case, when $\zeta$ has only one largest eigenvalue, then $\Psi
= |\Lambda(r_1)| = 1$ and so $C_q(L) - C_q(\infty) $ is dominated by a simple
exponential decay at large $L$.

For the case of multiple largest eigenvalues originating from the same cycle in
$\zeta$'s graph, then $\Psi = |\Lambda(r_1)|  > 1$. And so, the asymptotic
behavior of $C_q(L) - C_q(\infty) $ is dominated by a decaying \emph{pattern}
of length $|\Lambda(r_1)|$.

For example, the Lollipop processes have an
exponentially decaying pattern of length-$N$  that dominates $C_q(L) -
C_q(\infty) $ for $L > \nu_0 = M$:
\begin{align}
\Psi &= |\Lambda(r_1)| = N
  ~.
\end{align}
This periodic behavior is apparent in the semi-log plots of
Figs.~\ref{fig:Lollipop_slope} and \ref{fig:Lollipop_offset} and is especially
emphasized in Fig.~\ref{fig:Lollipop_period} which shows that $\Psi = N$ for
various $N$.  The figures demonstrate excellent agreement with our qualitative
expectations from the above approximations.

Showing the effect of different $\nu_0$, Fig.~\ref{fig:Lollipop_offset}
emphasizes that the initial rolloff of $C_q(L) - C_q(\infty)$ is due to $L \leq
\nu_0 = M$. The dominant asymptotic behavior is reached soon after $L = \nu_0$
in this case since the remaining (i.e., nonzero) eigenvalues of the QPMM
transition dynamic $\zeta$ are all in the largest-magnitude set $\Lambda(r_1)$.
In other words, Thm.~\ref{thm:DCqL}'s Eq. (\ref{eq:CqLAsymptoticBehavior}) is
not only approximated by but, in this case, also equal to the simpler
expression in Coro.~\ref{paul4}, since $r_2 = 0$.

The slope $r_1$ indicated in Figs.~\ref{fig:Lollipop_slope} and
\ref{fig:Lollipop_offset} corresponds to the asymptotic decay rate of the
envelope for $C_q(L) - C_q(\infty)$. This asymptotic decay rate is a function
of both $N$ and $p$, since for Lollipop:
\begin{align}
r_1 &= \bigl[ (1-p)(1-q) \bigr]^{1/N}
  ~.
\end{align}
Figure~\ref{fig:Lollipop_slope} shows that we have indeed identified the
correct slope for different $p$.

The central asymptotic features of $C_q(L) - C_q(\infty)$ are all captured
succinctly by Thm.~\ref{the:CqLAsymptoticBehavior}: First, the asymptotic
behavior of $C_q(L) - C_q(\infty)$ is exponentially decreasing at a rate of
$r_1$, which is the spectral radius of $\zeta$. Second, this exponential
envelope is modulated by an asymptotic $\Psi$-periodic structure, where $\Psi$
is the least common multiple of slowest-decaying QPMM cycle-lengths.

\section{Conclusion}

We developed a detailed analytical theory of how to optimally compress a
classical, stationary finite-memory stochastic process using a quantum channel.
This required introducing a new quantum state-machine representation (\qMLs),
carefully constructing its codewords and quantitatively monitoring their
overlaps (via the quantum pairwise-merger machine), and utilizing a new matrix
formulation of the overlap density matrix (abridged Gram matrix). Applying
spectral decomposition then lead directly to closed-form expressions for the
quantum coding costs at any codeword length, including infinite length.

The theoretical advances give an extremely efficient way to probe the behavior
of quantum compression, both analytically and, when symbolic calculation become
arduous, numerically. Analyzing selected example processes illustrated the
required calculations and also the range of phenomena to be expected when compressing memoryful processes.

Particular phenomena we reported for the first time here included (i) details
of how a process's cryptic order determines its quantum compressibility, (ii)
transient and persistent contributions to quantum compression costs, (iii)
exponential convergence to optimum compression, and (iv) oscillations in the
convergence that reveal how a process gives up its crypticity with increasing
codeword length. Results that hold for finite and infinite Markov and cryptic
order processes.

The overall result appears as a rather complete quantitative theory of the
informational properties of quantum channels used to compress classical
processes, including finite and infinite codeword closed-form expressions. That
said, many issues remain, both technical and philosophical. We believe,
however, that the approach's mathematical grounding and analytical and
numerical efficiency will go some distance to solving them in the near future.

For example, one of the abiding questions is the meaning of process crypticity
$\PC = \Cmu - \EE$---the difference between a process' predictable information
or excess entropy $\EE$ and its stored state information or statistical
complexity $\Cmu$ \cite{Crut08a,Crut08b}. Most directly, $\PC$ measures how
much state information ($\Cmu$) is hidden from observation ($\EE$). Cryptic
processes and even those with infinite cryptic order dominate the space of
classical processes \cite{Jame10a}. This means that generically we can compress
$\Cmu$ down to $C_q(L)$. However, this begs the question of what crypticity is
in the quantum domain. Now that we can work analytically in the infinite-length
limit, we can explore the \emph{quantum crypticity} $\PC_q = C_q (\infty) -
\EE$. From our studies, some not reported here, it appears that one cannot
compress the state information all the way down to the excess entropy. Why? Why
do not quantum models exist of ``size'' $\EE$ bits? Does this point to a
future, even more parsimonious physical theory? Or, to a fundamental limitation
of communication that even nature must endure, as it channels the past through
the present to the future?

For another, are we really justified in comparing Shannon bits ($\Cmu$) to
qubits ($C_q$)? This is certainly not a new or recent puzzle. However, the
results on compression bring it to the fore anew. And, whatever the outcome,
the answer will change our view of what physical pattern and structure are.
Likely, the answer will have a profound effect. Assuming the comparison is
valid, why is there a level of classical reality that is more structurally
complex when, as we demonstrated and now can calculate, processes are more
compactly represented quantum mechanically?


\acknowledgments

We thank Ryan James for helpful conversations. The authors thank the Santa Fe
Institute for its hospitality during visits. JPC is an SFI External Faculty
member. This material is based upon work supported by, or in part by, the John
Templeton Foundation and U. S. Army Research Laboratory and the U. S. Army
Research Office under contract W911NF-13-1-0390.

\appendix 

\section{Quantum Overlaps and Cryptic Order}
\label{APPOVER}

{\textbf{Lemma \ref{lem:Lmerge}.} 
Given an \eM\ with cryptic order $k$: for $L \leq k$, there exists an $L-$merge; for $ L > k$, there exists no $L-$merge.

{\ProLem 
By definition of cryptic order $k$:
\begin{align*}
 \H[\St_k | \MS_{0:\infty}]=0  ~,
\end{align*}
which means that for any given $\future$ there exists a unique $\st_k$. Since
$k$ is the minimum such length, for $L=k-1$ there exists some word
$\ms_{0:\infty}$ that leaves uncertainty in causal state $\St_{k-1}$. Call two
of these uncertain $\St_{k-1}$ states A and B ($A\neq B$). Tracing
$\ms_{0:\infty}$ backwards from A and B, we produce two state paths. These
state paths must be distinct at each step due to eM\ unifilarity---if they
were not distinct at some step, they would remain so for all states going
forward, particularly at $\St_{k-1}$. The next symbol $\ms_k$ must take A and B
to the same next state $F$ or violate the assumption of cryptic order $k$.
These two state paths and the word $\ms_{0:k}$ and the final state F make up a
$k$-merger, meaning that cryptic order $k$ implies the existence of a $k$-merger.

By removing states from the left side of this $k$-merger, it is easy to see
that a $k$-merger implies the existence of all shorter $L$-mergers.

By unifilarity again, $\H[\St_k | \Future] = 0 \rightarrow \H[\St_L | \Future]
= 0$, for all $L \geq k$. Assume there exists an $L$-merger for $L > k$ with
word $w$. By definition of $L$-merger, there is then uncertainty in the state
$\St_{L-1}$. This uncertainty exists for any word with $w$ as the prefix---a
set with nonzero probability. This contradicts the definition of cryptic order.}

{\textbf{Theorem \ref{thm:qq}.}
Given a process with cryptic order $k$, for each $L \in [0,k]$, each quantum
overlap $\braket{\eta_i(L)| \eta_j(L)} $ is a nondecreasing function of $L$.
Furthermore, for each $L \in [1,k]$, there exists at least one overlap that is
increased (as a result of a corresponding $L$-merge). For all remaining $L \geq
k$, each overlap takes a constant value $\braket{\eta_i(k)| \eta_j(k)}$. 

{\ProLem We directly calculate:
\begin{align*}
\braket{\eta_a(L)| \eta_b(L)} &= \!\!\!\!  \sum_{\substack{w , w' \in \MeasAlphabet^L \\ j_L , l_L \in \{ i \}_{i=1}^M }}\sqrt{T^{(w)}_{al_L}} \sqrt{T^{(w')}_{bj_L}} \braket{w | w'} \braket{\st_{l_L} | \st_{j_L}}
\\
&=\sum_{w,j_L}\sqrt{T^{(w)}_{aj_L}}\sqrt{T^{(w)}_{bj_L}}.
\end{align*}
So, we have:
\begin{align*}
&\braket{\eta_a (L+1)| \eta_b(L+1)}  \\
& \quad =\sum_{\substack{w^{\prime} \in \MeasAlphabet^{L+1}\\j_{L+1}}}\sqrt{T^{(w^{\prime})}_{aj_{L+1}}}\sqrt{T^{(w^{\prime})}_{bj_{L+1}}}\\
& \quad =\sum_{\substack{w \in \MeasAlphabet^{L},s \in \MeasAlphabet\\j_L,l_L,j_{L+1}}}\sqrt{T^{(w)}_{aj_{n}}}\sqrt{T^{(s)}_{j_nj_{L+1}}}\sqrt{T^{(w)}_{bl_{L}}}\sqrt{T^{(s)}_{l_Lj_{L+1}}}\\
& \quad =\sum_{\substack{w \in \MeasAlphabet^{L},s \in \MeasAlphabet\\j_L,j_{L+1}}}\sqrt{T^{(w)}_{aj_{L}}}\sqrt{T^{(s)}_{j_Lj_{L+1}}}\sqrt{T^{(w)}_{bj_{L}}}\sqrt{T^{(s)}_{j_Lj_{L+1}}}\\
& \quad \quad  +\sum_{\substack{w \in \MeasAlphabet^{L},s \in \MeasAlphabet \\j_L \neq l_L,j_{L+1}}}\sqrt{T^{(w)}_{aj_{L}}}\sqrt{T^{(s)}_{j_Lj_{L+1}}}\sqrt{T^{(w)}_{bl_{L}}}\sqrt{T^{(s)}_{l_Lj_{L+1}}}
  ~,
\end{align*}	
The first sum represents the overlaps obtained already at length $L$.
To see this, we split the sum to two parts, where the first contains:
\begin{align*}
\sum_{\substack{w \in \MeasAlphabet^{L},s \in \MeasAlphabet\\j_L,j_{L+1}}}
  & \sqrt{T^{(w)}_{aj_{L}}}
  \sqrt{T^{(s)}_{j_Lj_{L+1}}}
  \sqrt{T^{(w)}_{bj_{L}}}
  \sqrt{T^{(s)}_{j_Lj_{L+1}}}\\
& =\sum_{\substack{w \in \MeasAlphabet^{L}\\j_L}}
  \sqrt{T^{(w)}_{aj_{L}}}
  \sqrt{T^{(w)}_{bj_{L}}}
  \Bigl( \sum_{\substack{s \in \MeasAlphabet\\j_{L+1}}}
  \sqrt{T^{(s)}_{j_Lj_{L+1}}}
  \sqrt{T^{(s)}_{j_Lj_{L+1}}} \, \Bigr) \\
& =\sum_{\substack{w\in \MeasAlphabet^{L}\\j_L}}
  \sqrt{T^{(w)}_{aj_L}}
  \sqrt{T^{(w)}_{bj_L}} \\
& =\braket{\eta_a(L)| \eta_b(L)}
  ~.
\end{align*}
We use Lemma \ref{lem:Lmerge} to analyze the second sum, which represents the change in the overlaps, finding that:
\begin{align*}
\sum_{\substack{w \in \MeasAlphabet^{L},s \in \MeasAlphabet\\j_L \neq l_L,j_{L+1}}}
  \sqrt{T^{(w)}_{aj_{L}}}
  \sqrt{T^{(s)}_{j_Lj_{L+1}}}
  \sqrt{T^{(w)}_{bl_{L}}}
  \sqrt{T^{(s)}_{l_Lj_{L+1}}} \geq 0
  ~,
\end{align*}
with equality when $L \geq k$. Summarizing:
\begin{align*}
\braket{\eta_a(L+1)| \eta_b(L+1)} \geq \braket{\eta_a(L)| \eta_b(L)}
  ~,
\end{align*}
with equality for $L \geq k$.
}

Note that while the set of overlaps continues to be augmented at each length up until the cryptic order, we do not currently have a corresponding statement about the nontrivial change in $C_q(L)$ or its monotonicity.

\section{Matrices and Their Entropy}
\label{DMGAG}

\subsection{Density Matrix}

The density matrix can now be expressed using a fixed $|\SSet|$-by-$|\SSet|$
matrix, valid for all $L$. Using the Gram-Schmidt procedure one can choose a new orthonormal basis. Let:
\begin{align*}
\ket{\eta_1(L)}&= \ket{ e_1^{(L)} } \\
\ket{\eta_2(L)}&= a_{21}^{(L)} \ket{ e_1^{(L)} } + a_{22}^{(L)} \ket{ e_2^{(L)} } \\
\ket{\eta_3(L)}&= a_{31}^{(L)} \ket{ e_1^{(L)} } + a_{32}^{(L)} \ket{ e_2^{(L)} } + a_{33}^{(L)} \ket{ e_3^{(L)} } \\
&\vdots
\end{align*}
and so on. Then:
\begin{align*}
a_{21}^{(L)} & =\braket{\eta_1(L)| \eta_2(L)} \\
& = \bra{ (\st_1, \st_2) } \bigl( \sum_{n=0}^L \zeta^n \bigr) \ket{ \synk } ~, \\ 
a_{22}^{(L)} &= \left( 1 - | \braket{\eta_1(L)| \eta_2(L)} |^2 \right)^{1/2} ~, \\
a_{31}^{(L)} & =\braket{\eta_1(L)| \eta_3(L)} \\
  & = \bra{ (\st_1, \st_3)} \bigl( \sum_{n=0}^L \zeta^n \bigr) \ket{ \synk }
  ~,
\end{align*}
and so on. Now it is useful to rewrite what we can in matrix form:
\begin{align*}
\begin{bmatrix}
\bra{\eta_1(L)} \\
\bra{\eta_2(L)} \\
\bra{\eta_3(L)} \\
\vdots \\
\bra{\eta_{|\SSet|}(L)} 
\end{bmatrix} 
= 
\underbrace{
  \begin{bmatrix}
  1 &  &  &   & 0 \\
  a_{21}^{(L)} & a_{22}^{(L)}  &  &  &  \\
  a_{31}^{(L)} & a_{32}^{(L)}  & a_{33}^{(L)} &   &  \\
  \vdots &  &  &  \ddots   &  \\
  a_{ |\SSet| 1}^{(L)} &  &  \cdots  &    &  a_{ |\SSet|  |\SSet|}^{(L)}
  \end{bmatrix}
}_{\equiv A_L} 
\, 
\begin{bmatrix}
\bra{ e_1^{(L)} } \\
\bra{ e_2^{(L)} } \\
\bra{ e_3^{(L)} } \\
\vdots \\
\bra{ e_{ |\SSet| }^{(L)} } 
\end{bmatrix} 
~,
\end{align*}
which defines the lower-triangular matrix $A_L$. Note that the rightmost matrix
of orthonormal basis vectors is simply the identity matrix, since we are
working in that basis.

In this new basis, we construct the $|\SSet|$-by-$|\SSet|$ density matrix as: 
\begin{align*}
\rho(L) 
&= \sum_{i=1}^{ |\SSet| }{\pi_i \ket{\eta_i(L)} \bra{\eta_i(L)}}\\
&= 
\begin{bmatrix}
\ket{\eta_1(L)} &
\cdots &
\ket{\eta_{ |\SSet| }(L)}
\end{bmatrix} 
\,
\underbrace{
\begin{bmatrix}
  \pi_1 & &  0 \\
  &  \ddots & \\
  0 & &  \pi_{ |\SSet| }
  \end{bmatrix}
}_{\equiv D_\pi}
\,
\begin{bmatrix}
\bra{\eta_1(L)} \\
\bra{\eta_2(L)} \\
\bra{\eta_3(L)} \\
\vdots \\
\bra{\eta_{ |\SSet| }(L)} 
\end{bmatrix} 
\\ 
& = A_L^\dagger D_\pi A_L
  ~.
\end{align*}

Since all entries are real, the conjugate transpose is the transpose. This
more general framework may be useful, however, if we want to consider the
effect of adding phase to the quantum states.

\subsection{von Neumann Entropy}

The quantum coding cost is:
\begin{align*}
C_q(L) & = -\text{tr} \left[ \rho(L)  \log \rho(L) \right] \\
  & = -\text{tr}
  \left[ A_L^\dagger D_\pi A_L \log (A_L^\dagger D_\pi A_L) \right] \\
  & = \, - \! \! \sum_{\lambda \in \Lambda_{A_L^\dagger D_\pi A_L}}
  \lambda \log \lambda ~.
\end{align*}
This is relatively easy to evaluate since the density matrix $\rho(L)$ is only
a $|\SSet|$-by-$|\SSet|$ function of $L$. Thus, we calculate $C_q(L)$
analytically from $\rho$'s spectrum. This, in a curious way, was already folded
into $\zeta$'s spectrum.

\subsection{Gram Matrix}

The $A_L$ matrix is burdensome due to nonlinear dependence on the overlap of
the quantum states. We show how to avoid this nonlinearity and instead obtain
the von Neumann entropy from a transformation that yields a linear relationship
with overlaps.

The \emph{Gram matrix}, with elements $G_{mn}^{(L)} = \sqrt{\pi_m \pi_n}
\braket{\eta_m(L)| \eta_n(L)}$, can be used instead of $\rho(L)$ to evaluate
the von Neumann entropy \cite{Jozsa00a}. In particular, $G^{(L)}$ has the same
spectrum as $\rho(L)$, even with the same multiplicities: $\Lambda_{ G^{(L)} }
= \Lambda_{\rho(L)}$, while $a_\lambda$, $g_\lambda$, and $\nu_\lambda$ remain
unchanged for all $\lambda$ in the spectrum. (This is a slightly stronger
statement than in Ref. \cite{Jozsa00a}, but is justified since $\rho(L)$ and
$G^{(L)}$ are both $|\SSet|$-by-$|\SSet|$ dimensional.)

Here, we briefly explore the relationship between $ \rho(L)$ and $G^{(L)}$ and,
then, focus on the closed-form expression for $G^{(L)}$. The result is more
elegant than $\rho(L)$, allowing us to calculate and understand $C_q(L)$ more
easily.

Earlier, we found that the density matrix can be written as:
\begin{align*}
\rho(L) &= A_L^\dagger D_\pi A_L ~,
\intertext{which can be rewritten as: }
\rho(L)&= 
A_L^\dagger D_\pi^{1/2} D_\pi^{1/2} A_L \\
&= \left( D_\pi^{1/2} A_L \right)^\dagger D_\pi^{1/2} A_L ~.
\end{align*}

It is easy to show that:
\begin{align*}
\text{tr} \left[  \bigl( D_\pi^{1/2} A_L \bigr)^\dagger D_\pi^{1/2} A_L \right] 
&=
\text{tr} \left[ D_\pi^{1/2} A_L  \bigl( D_\pi^{1/2} A_L \bigr)^\dagger \right] \\ 
&=
\text{tr} \left[ D_\pi^{1/2} A_L   A_L^\dagger D_\pi^{1/2} \right]
  ~.
\end{align*}
This means that the sum of the eigenvalues is conserved in transforming from
$A_L^\dagger D_\pi A_L$ to $D_\pi^{1/2} A_L   A_L^\dagger D_\pi^{1/2}$.  It is
less obvious that the spectrum is also conserved, but this is also true, and
even easy to prove. (Observe that $AB \vec{v} = \lambda \vec{v} \Longrightarrow
BAB \vec{v} = \lambda B \vec{v} \Longrightarrow BA(B \vec{v}) = \lambda (B
\vec{v})$.) Interestingly, the new object turns out to be exactly the Gram
matrix, which was previously introduced, although without this explicit
relationship to the density matrix. We now see that:
\begin{align*}
&D_\pi^{1/2}  A_L   A_L^\dagger D_\pi^{1/2} \\
&= 
D_\pi^{1/2} 
\,
\begin{bmatrix}
\bra{\eta_1(L)} \\
\vdots \\
\bra{\eta_{|\SSet|}(L)} 
\end{bmatrix} 
\,
\begin{bmatrix}
\ket{\eta_1(L)}&
\cdots &
\ket{\eta_{|\SSet|}(L)}  
\end{bmatrix} 
\,
D_\pi^{1/2} 
\\
&= 
\begin{bmatrix}
\sqrt{\pi_1} \bra{\eta_1(L)}  \\
\vdots \\
\sqrt{\pi_{|\SSet|}} \bra{\eta_{|\SSet|}(L)} 
\end{bmatrix} 
\,
\begin{bmatrix}
\sqrt{\pi_1} \ket{\eta_1(L)}&
\cdots &
\sqrt{\pi_{|\SSet|}} \ket{\eta_{|\SSet|}(L)}
\end{bmatrix} 
\\
&= 
\begin{bmatrix}
\sqrt{\pi_1 \pi_1} \braket{\eta_1(L)| \eta_1(L)} & \cdots & \sqrt{\pi_1 \pi_{|\SSet|}} \braket{\eta_1(L)| \eta_{|\SSet|}(L)}\\
\vdots & \ddots & \vdots \\
\sqrt{\pi_{|\SSet|} \pi_1} \braket{\eta_{|\SSet|}(L)| \eta_1(L)}& \cdots &  \sqrt{\pi_1 \pi_{|\SSet|}} \braket{\eta_{|\SSet|}(L)| \eta_{|\SSet|}(L)}
\end{bmatrix} 
\\
& = G^{(L)}
  ~.
\end{align*}

Since the spectrum is preserved, we can use the Gram matrix directly to compute the von Neumann entropy:
\begin{align*}
C_q(L) 
&= \, - \!  \sum_{\lambda \in \Lambda_{G^{(L)} } } \lambda \log \lambda \\
&= -\text{tr} \left[ G^{(L)} \log G^{(L)} \right]
  ~.
\end{align*}

\subsection{Abridged-Gram Matrix}

Transforming to the Gram matrix suggests a similar and even more helpful
simplification that can be made while preserving the spectrum.
Define the \emph{abridged Gram matrix} to be:
\begin{align*}
\widetilde{G}^{(L)} & \equiv D_\pi A_L A_L^\dagger \nonumber \\
& = D_\pi
\,
\begin{bmatrix}
\bra{\eta_{1}(L)} \\
\vdots \\
\bra{\eta_{|\SSet|}(L)} 
\end{bmatrix} 
\,
\begin{bmatrix}
\ket{\eta_{1}(L)}&
\cdots &
\ket{\eta_{|\SSet|}(L)}  
\end{bmatrix} 
\\
& = 
\begin{bmatrix}
\pi_1 \braket{\eta_1(L)| \eta_1(L)} & \cdots & \pi_1\braket{\eta_1(L)| \eta_{|\SSet|}(L)} \\
\vdots & \ddots & \vdots \\
\pi_{|\SSet|} \braket{\eta_{|\SSet|}(L)| \eta_1(L)} & \cdots &  \pi_{|\SSet|} \braket{\eta_{|\SSet|}(L)| \eta_{|\SSet|}(L)} 
\end{bmatrix} 
\end{align*}
Clearly, this preserves the same trace as the density matrix and previous Gram
matrix. It also preserves the spectrum. And, it has the advantage of not using
square-roots of two different state probabilities in each element. Rather it
has a single probability attached to each element.

Since the spectrum is preserved, we can use the abridged-Gram matrix to compute
the von Neumann entropy:
\begin{align}
C_q(L) 
& = \, - \!  \sum_{\lambda \in \Lambda_{\widetilde{G}^{(L)} } }
  \lambda \log \lambda \\
&= -\text{tr} \left[ \widetilde{G}^{(L)} \log \widetilde{G}^{(L)} \right]
  ~.
\end{align}

\section{Examples}
\label{appex}

Exploring several more examples will help to illustrate the methods and
lead to additional observations.

\subsection{Biased Coins Process}

The Biased Coins Process provides a first, simple case that realizes a
nontrivial quantum state entropy \cite{Gu12a}. There are two biased coins,
named $A$ and $B$. The first generates $1$ with probability $q$; the second,
$0$ with probability $p$. A coin is picked and flipped, generating outputs $0$
or $1$. With probability $q$ the other coin is used next similarly with
different probability. Its two causal-state $\eM$ is shown in Fig. \ref{fig:PC_eM}.

\begin{figure}
\advance\leftskip 0.8cm
\includegraphics[width=1.5\linewidth]{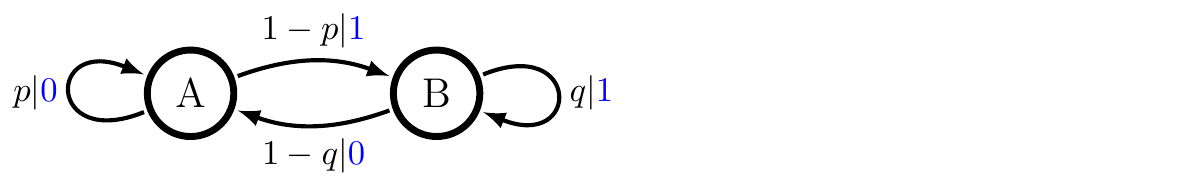}
  \caption{\EM\ for the Biased Coins Process.}
\label{fig:PC_eM}
\end{figure}

\begin{figure}
\advance\leftskip 1.7cm
\includegraphics[width=1.5\linewidth]{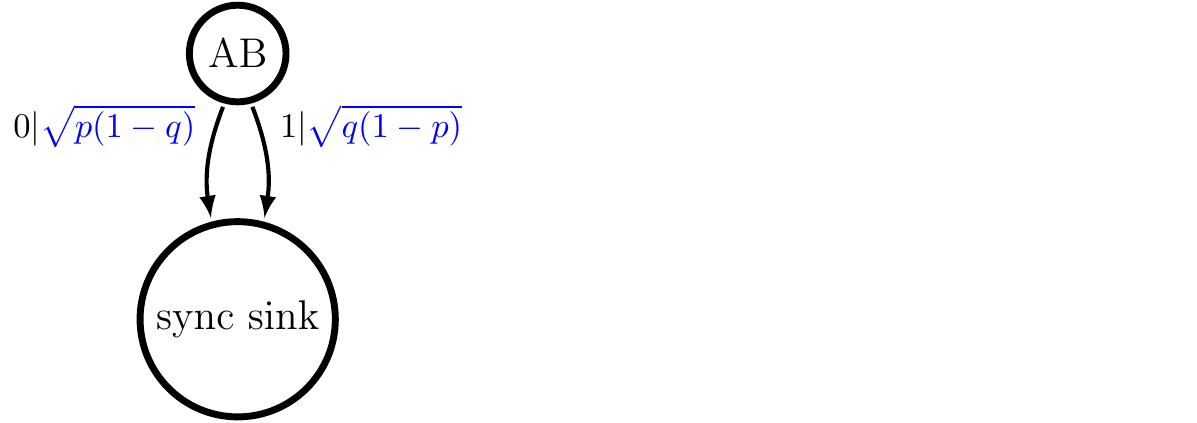}
\caption{QPMM for the Biased Coins Process.}
\label{fig:PC_QPMM}
\end{figure}

After constructing the QPMM for the Biased Coins Process, as outlined in Figs.
\ref{fig:PC_eM} and \ref{fig:PC_QPMM}, we observe:
\begin{align*}
\zeta^{(0)} &= 
\begin{bmatrix}
0 & \sqrt{p(1-q)} \\
0 & 0 
\end{bmatrix} ~, \\
\zeta^{(1)} &= 
\begin{bmatrix}
0 & \sqrt{q(1-p)} \\
0 & 0 
\end{bmatrix} ~,
\intertext{and so:}
\zeta &= 
\begin{bmatrix}
0 & \beta \\
0 & 0 
\end{bmatrix}
  ~,
\end{align*}
where we defined $\beta \equiv \sqrt{p(1-q)} + \sqrt{q(1-p)}$. Let's also
define the suggestive quantity $\gamma \equiv (1-\beta^2)^{-1/2}$.

The only overlap to consider is 
$\braket{\eta_{A}(L)| \eta_{B}(L)}$.  For this, we note that 
$\bra{(A,B)}  = \begin{bmatrix} 1 & 0 \end{bmatrix}$.
Also, 
$\ket{\synk } = \begin{bmatrix} 0 & 1 \end{bmatrix}^{\top}$.

Spectrally, $\zeta$ here is a nilpotent matrix with only a zero eigenvalue with
index two: $\Lambda_\zeta = \{ 0 \}$ and $\nu_0 = 2$. Since the projection
operators must sum to the identity, we have $\zeta_0 = I$.

$\zeta^L$ is the null matrix for $L > 1$, so either by
Eq.~\eqref{eq:TruncatedOverlap_wDepth} or by Eq.~\eqref{eq:TruncatedOverlap},
we have:
\begin{align*}
\braket{\eta_{A}(L)| \eta_{B}(L)}&= 
\sum_{m=1}^{\min \{ L, \, 1 \} }  \bra{(A,B)} \zeta^{m}  \ket{\synk }
  ~.
\end{align*}
That is:
\begin{align*}
\braket{\eta_{A}(L)| \eta_{B}(L)}&= 
\begin{cases}
0 & \text{if } L=0 ~,\\
\beta & \text{if } L \geq 1 ~.
\end{cases}
\end{align*}

\subsubsection{Entropy from the Density Matrix}

For the density matrix, we turn to the $L$-dependent orthonormal basis $\{
\ket{ e_1^{(L)} } , \ket{ e_2^{(L)} } \}$ and use the stationary distribution
over $\SSet$: $\pi = \begin{bmatrix} p/(p+q) & q/(p+q) \end{bmatrix}$.

Apparently, for $L=0$ we have: $\ket{\eta_{A}(0)} = \ket{ e_1^{(0)} } $ and
$\ket{\eta_{B}(0)}= \ket{ e_2^{(0)} }$. Hence, $\rho(0)= D_\pi $ and $C_q(0)=
H_2(p/(p+q))=\Cmu$ qubits.

For $L \geq 1$ we have: $\ket{\eta_{A}(L)} = \ket{ e_1^{(L)} }$ and $\ket{\eta_{A}(L)} = a_{21}^{(L)} \ket{ e_1^{(L)} } + a_{22}^{(L)} \ket{ e_2^{(L)} }$, where $a_{21}^{(L)} = \braket{\eta_{A}(L)| \eta_{B}(L)}= \beta$ and $a_{22}^{(L)} = (1- \beta^2)^{1/2} = \gamma^{-1}$ for $L \geq 1$.
We find that:
\begin{align*}
A_L &= 
\begin{bmatrix}
1 & 0 \\
\beta & \gamma^{-1}
\end{bmatrix} ~,
~\text{for}~ L \geq 1 ~.
\end{align*}
Hence, for $L \geq 1$ the density matrix is:
\begin{align*}
\rho(L)&= 
A_L^\dagger D_\pi A_L \\ 
&= 
\begin{bmatrix}
1 & \beta \\
0 & \gamma^{-1}
\end{bmatrix}
\,
\begin{bmatrix}
\tfrac{p}{p+q} & 0 \\
0 & \tfrac{q}{p+q}
\end{bmatrix}
\,
\begin{bmatrix}
1 & 0 \\
\beta & \gamma^{-1}
\end{bmatrix}
\\
&= 
\tfrac{1}{p+q}
\begin{bmatrix}
p & q \beta \\
0 & q \gamma^{-1}
\end{bmatrix}
\,
\begin{bmatrix}
1 & 0 \\
\beta & \gamma^{-1}
\end{bmatrix}
\\
&= 
\tfrac{q}{p+q}
\begin{bmatrix}
\tfrac{p}{q}+\beta^2 & \beta / \gamma \\
\beta / \gamma & 1- \beta^2
\end{bmatrix} 
  ~.
\end{align*}

Since:
\begin{align*}
\det (\rho(L) - \lambda I)
  = \lambda^2 - \lambda + \tfrac{pq}{(p+q)^2}(1-\beta^2)
  ~,
\end{align*}
we find $\rho(L)$'s eigenvalues to be:
\begin{align*}
\Lambda_{\rho(L)} = \bigl\{ \tfrac{1}{2} \pm \tfrac{1}{2(p+q)} \sqrt{ 4pq \beta^2 + (p-q)^2 } \, \bigr\}
~,
\end{align*}
which yields the von Neumann entropy for $L \geq 1$:
\begin{align*}
C_q(L) & = - \sum_{\lambda \in \Lambda_{\rho(L)}} \lambda \log \lambda 
  ~.
\end{align*}

\subsubsection{Entropy from the Abridged-Gram Matrix}

The abridged-Gram matrix for the Biased Coins Process is:
\begin{align*}
\widetilde{G}^{(L)}
&= 
D_\pi
\,
\begin{bmatrix}
\braket{\eta_{A}(L)| \eta_{A}(L)}&\braket{\eta_{A}(L)| \eta_{B}(L)}\\
\braket{\eta_{B}(L)| \eta_{A}(L)}& \braket{\eta_{B}(L)| \eta_{B}(L)}  
\end{bmatrix} 
~.
\end{align*}
Specifically, we have for $L \geq 1$:
\begin{align*}
\widetilde{G}^{(0)}
& =
\tfrac{1}{p+q}
\,
\begin{bmatrix}
p & 0 \\
0 & q
\end{bmatrix} 
\,
\begin{bmatrix}
1 & 0 \\
0 & 1
\end{bmatrix} \\
& = 
\tfrac{1}{p+q}
\,
\begin{bmatrix}
p & 0 \\
0 & q
\end{bmatrix} 
\intertext{and:}
\widetilde{G}^{(L)}
& =
\tfrac{1}{p+q}
\,
\begin{bmatrix}
p & 0 \\
0 & q
\end{bmatrix} 
\,
\begin{bmatrix}
1 & \beta \\
\beta & 1
\end{bmatrix} \\
& = 
\tfrac{1}{p+q}
\,
\begin{bmatrix}
p & p \beta \\
q \beta & q
\end{bmatrix} 
~.
\end{align*}

$\widetilde{G}^{(0)}$'s eigenvalues are simply its diagonal entries. So, 
$C_q(0)= H_2(p / (p+q))$ qubits.
For $L \geq 1$:
\begin{align*}
\det (\widetilde{G}^{(L)} - \lambda I)
  = \lambda^2 - \lambda + \tfrac{pq}{(p+q)^2}(1-\beta^2)
  ~, 
\end{align*}
which gives the same value for eigenvalues and entropy.

As the new method illustrates, there is no need to construct the density matrix.
Instead, one uses the abridged-Gram matrix, which can be easily calculated from
quantum overlaps. Clearly, the abridged-Gram matrix method is more elegant for
our purposes. This is evident even at $|\SSet|=2$. This is even more critical
for more complex processes since $A_L$  grows as $|\SSet|$ grows.

\subsection{($R$--$k$)-Golden Mean Process}

\begin{figure}
\advance\leftskip 1.1cm
 \caption{\EM\ for the ($R$--$k$)-Golden Mean Process.}
\label{fig:RKGM_eM}
\end{figure}

The ($R$--$k$)-Golden Mean Process is constructed to have Markov-order $R$ and
cryptic-order $k$. Its \eM\ is shown in Fig.~\ref{fig:RKGM_eM}. The
$0^\text{th}$ state $\st_0$ has probability $\pi_0 = 1 / (R+k-p(R+k-1))$ while
all other states $\st_i$ have probability $\pi_i = (1-p) \pi_0$.

Its QPMM is strictly tree-like with depth $d = k+1$ and maximal width $k$.  All
edges have a unit weight except for those edges leaving $A$-paired states. The
latter edges, numbering $k$ in total, have an associated weight of $\sqrt{p}$.

The eigenvalues of the abridged-Gram matrix can be obtained from:
\begin{align*}
\det (\widetilde{G}^{(L)} & - \lambda I ) 
= (\pi_1-\lambda)^{R+k-\min(L,k) - 1} \\
& \times \left| 
\begin{matrix}
\pi_0 - \lambda & \pi_0 \sqrt{p} & \cdots & \pi_0 \sqrt{p}^{\min(L,k)} \\
\pi_1 \sqrt{p}  & \pi_1 - \lambda &  & \pi_1 \sqrt{p}^{\min(L,k)-1} \\
\vdots &   &  \ddots & \\ 
\pi_1 \sqrt{p}^{\min(L,k)} & & & \pi_1 - \lambda
\end{matrix} 
\right| \\
& \quad\quad\quad = 0
  ~,
\end{align*} 
which directly yields the von Neumann entropy. Note that although the $C_q(L)$
is \emph{not} actually linear in $L$, it appears approximately linear.

We observe that $\boldsymbol{\pi}$ is invariant under the simultaneous change
of:
\begin{align}
R' = R + m ~, \text{ while } \; k' = k-m
  ~,
\label{eq:RkTransformation}
\end{align}
for any $m \in \mathbb{Z}$.
Although we insist on maintaining $R' \geq k' \geq 0$ for preservation of their functional roles.
Furthermore, this transformation preserves $\det (\widetilde{G}^{(L)} - \lambda I )$
for $L \leq k$ and $L \leq k'$.  Hence $C_q(L)$ is invariant to the simultaneous 
transformation of Eq.~\eqref{eq:RkTransformation} for $L \leq k$ and $L \leq k'$,
which explains our near-linear observation in Fig.~\ref{fig:RkGM_Cq}'s caption.

To give an explicit example, let's consider the ($4$--$3$)-GM Process of Fig.
\ref{fig:43GM_eM}. State $A$ has probability $\pi_A = 1 / (7-6p)$ while all
other states have probability $\pi_i = (1-p)/(7-6p)$. Let's calculate:
\begin{itemize}
\item For $L=0$:
\begin{align*}
\det (\widetilde{G}^{(0)} - \lambda I ) = (\pi_B - \lambda)^6 (\pi_A - \lambda) ~,
\end{align*}
yielding $\Lambda_{\widetilde{G}^{(0)}} = \{ \pi_B, \pi_A \}$ (with $a_{\pi_B} = 6$) and 
\begin{align*}
C_q{(0)} = -6 \pi_B \log \pi_B - \pi_A \log \pi_A ~.
\end{align*}
\item For $L=1$:
\begin{align*}
& \det (\widetilde{G}^{(1)} - \lambda I ) \\
& \quad = (\pi_B - \lambda)^5
\left[ \lambda^2 - (\pi_A + \pi_B) \lambda + \pi_A \pi_B (1-p) \right] ~,
\end{align*}
yielding 
$\Lambda_{\widetilde{G}^{(1)}} = \{ \pi_B, c_+, c_-  \}$ with 
$c_\pm = \tfrac{1}{2} (\pi_A + \pi_B) \pm \tfrac{1}{2} \left[ (\pi_A + \pi_B)^2 -4 \pi_A \pi_B (1-p) \right]^{1/2}$ 
(and with $a_{\pi_B} = 5$), and:
\begin{align*}
C_q{(1)} = -5 \pi_B \log \pi_B - c_+ \log c_+ - c_- \log c_- ~.
\end{align*}
\item For $L = 2$:
\begin{align*}
& \det (\widetilde{G}^{(2)} - \lambda I ) \\
& \quad = (\pi_B - \lambda)^4
\left| 
\begin{matrix}
\pi_A - \lambda & \pi_A p^{1/2} & \pi_A p \\
\pi_B p^{1/2} & \pi_B - \lambda & \pi_B p^{1/2}    \\
\pi_B p & \pi_B p^{1/2}   &  \pi_B - \lambda 
\end{matrix} 
\right| ~.
\end{align*} 
\item For $L \geq 3$:
\begin{align*}
& \det (\widetilde{G}^{(L)} - \lambda I ) \\
  & \quad = \det (\widetilde{G}^{(3)} - \lambda I ) \\
  & \quad = (\pi_B - \lambda)^3
\left| 
\begin{matrix}
\pi_A - \lambda & \pi_A p^{1/2} & \pi_A p & \pi_A p^{3/2} \\
\pi_B p^{1/2} & \pi_B - \lambda & \pi_B p^{1/2}   & \pi_B p \\
\pi_B p & \pi_B p^{1/2}   &  \pi_B - \lambda & \pi_B p^{1/2} \\ 
\pi_B p^{3/2} & \pi_B p & \pi_B p^{1/2} & \pi_B - \lambda
\end{matrix} 
\right|
~.
\end{align*} 
\end{itemize}


\bibliography{chaos}

\begin{thebibliography}{10}

\bibitem{Crut12a}
J.~P. Crutchfield.
\newblock Between order and chaos.
\newblock {\em Nature Physics}, 8(January):17--24, 2012.

\bibitem{Ball99a}
P.~Ball.
\newblock {\em The Self-Made Tapestry: Pattern Formation in Nature}.
\newblock Oxford University Press, New York, 1999.

\bibitem{Hoyl06a}
R.~Hoyle.
\newblock {\em Pattern Formation: {An} Introduction to Methods}.
\newblock Cambridge University Press, New York, 2006.

\bibitem{Kant06a}
H.~Kantz and T.~Schreiber.
\newblock {\em Nonlinear Time Series Analysis}.
\newblock Cambridge University Press, Cambridge, United Kingdom, second
  edition, 2006.

\bibitem{Crut88a}
J.~P. Crutchfield and K.~Young.
\newblock Inferring statistical complexity.
\newblock {\em Phys. Rev. Let.}, 63:105--108, 1989.

\bibitem{Shal98a}
C.~R. Shalizi and J.~P. Crutchfield.
\newblock Computational mechanics: Pattern and prediction, structure and
  simplicity.
\newblock {\em J. Stat. Phys.}, 104:817--879, 2001.

\bibitem{Gu12a}
M.~Gu, K.~Wiesner, E.~Rieper, and V.~Vedral.
\newblock Quantum mechanics can reduce the complexity of classical models.
\newblock {\em Nature Comm.}, 3(762):1--5, 2012.

\bibitem{Gmein11a}
P.~Gmeiner.
\newblock Equality conditions for internal entropies of certain classical and
  quantum models.
\newblock {\em arXiv:1108.5303}, 2011.

\bibitem{Maho15a}
J.~R. Mahoney, C.~Aghamohammadi, and J.~P. Crutchfield.
\newblock Occam's quantum strop: {Synchronizing} and compressing classical
  cryptic processes via a quantum channel.
\newblock 2015.
\newblock Santa Fe Institute Working Paper 15-08-030; arxiv.org:1508.02760
  [quant-ph].

\bibitem{Crut08a}
J.~P. Crutchfield, C.~J. Ellison, and J.~R. Mahoney.
\newblock Time's barbed arrow: {Irreversibility}, crypticity, and stored
  information.
\newblock {\em Phys. Rev. Lett.}, 103(9):094101, 2009.

\bibitem{Ash65a}
R.~B. Ash.
\newblock {\em Information Theory}.
\newblock John Wiley and Sons, New York, 1965.

\bibitem{Jame10a}
R.~G. James, J.~R. Mahoney, C.~J. Ellison, and J.~P. Crutchfield.
\newblock Many roads to synchrony: Natural time scales and their algorithms.
\newblock {\em Physical Review E}, 89:042135, 2014.

\bibitem{Crut10a}
J.~P. Crutchfield, C.~J. Ellison, J.~R. Mahoney, and R.~G. James.
\newblock Synchronization and control in intrinsic and designed computation:
  {An} information-theoretic analysis of competing models of stochastic
  computation.
\newblock {\em CHAOS}, 20(3):037105, 2010.

\bibitem{Maho09a}
J.~R. Mahoney, C.~J. Ellison, and J.~P. Crutchfield.
\newblock Information accessibility and cryptic processes.
\newblock {\em J. Phys. A: Math. Theo.}, 42:362002, 2009.

\bibitem{Maho11a}
J.~R. Mahoney, C.~J. Ellison, R.~G. James, and J.~P. Crutchfield.
\newblock How hidden are hidden processes? a primer on crypticity and entropy
  convergence.
\newblock {\em CHAOS}, 21(3):037112, 2011.

\bibitem{Franklin00}
J.~N. Franklin.
\newblock {\em Matrix Theory}.
\newblock Dover Publications, New York, 2000.

\bibitem{Dunford54}
N.~Dunford.
\newblock Spectral operators.
\newblock {\em Pacific J. Math.}, 4(3):321--354, 1954.

\bibitem{Crut13a}
J.~P. Crutchfield, P.~M. Riechers, and C.~J. Ellison.
\newblock Exact complexity: {Spectral} decomposition of intrinsic computation.
\newblock 2013.
\newblock Santa Fe Institute Working Paper 13-09-028; arXiv:1309.3792
  [cond-mat.stat-mech].

\bibitem{Yosida95}
K.~Yosida.
\newblock {\em Functional Analysis}.
\newblock Classics in Mathematics. Cambridge University Press, 1995.

\bibitem{Jozsa00a}
R.~Jozsa and J.~Schlienz.
\newblock Distinguishability of states and von {Neumann} entropy.
\newblock {\em Phys. Rev. A}, 62:012301, 2000.

\bibitem{Note1}
In principle, we need to consider two cases: the pattern decays to $C_q(\infty
  )$ from above or from below. In either case, the decay of the $C_q(L) -
  C_q(\infty )$ pattern would be exponential. However, it is known that
  $C_q(L)$ is strictly less than $C_\mu = C_q(0)$ for any $L$ for any
  noncounifilar process (and equal otherwise). Hence, we expect that $C_q(L)$
  always decays from above, as strongly corroborated by extensive numerical
  exploration.

\bibitem{Crut08b}
C.~J. Ellison, J.~R. Mahoney, and J.~P. Crutchfield.
\newblock Prediction, retrodiction, and the amount of information stored in the
  present.
\newblock {\em J. Stat. Phys.}, 136(6):1005--1034, 2009.

\end{thebibliography}

\end{document}